\documentclass[mnsc,nonblindrev]{informs3}
\usepackage{arydshln}
\usepackage{algorithm,algorithmic}
\usepackage{array,multirow}
\usepackage{caption}
\usepackage{subcaption}

\usepackage[normalem]{ulem}

\OneAndAHalfSpacedXI 

\usepackage{natbib}
 \bibpunct[, ]{(}{)}{,}{a}{}{,}%
 \def\bibfont{\small}%
 \def\BIBand{and}%

\TheoremsNumberedThrough     
\ECRepeatTheorems

\EquationsNumberedThrough    

\usepackage{graphicx}
\usepackage{tabularx}

\usepackage{mwe}

\usepackage{calc}

\usepackage{accents}

\begin{document}


\RUNAUTHOR{Chopra, Qiu, and Shim}

\RUNTITLE{Parallel Power System Restoration}

\TITLE{Parallel Power System Restoration}

\ARTICLEAUTHORS{%
\AUTHOR{Sunil Chopra\footnote{The authors are listed in alphabetical order and have contributed equally. All authors are first authors.}}
\AFF{Kellogg School of Management, Northwestern University, Evanston, IL 60208, USA, \EMAIL{s-chopra@kellogg.northwestern.edu}} 
\AUTHOR{Feng Qiu*}
\AFF{Energy Systems Division, Argonne National Laboratory, Lemont, IL 60439, USA, \EMAIL{fqiu@anl.gov}}
\AUTHOR{Sangho Shim*}
\AFF{School of Engineering, Mathematics and Science, Robert Morris University, Moon Twp, PA 15108, USA, \EMAIL{shim@rmu.edu}} 
} 

\ABSTRACT{
\noindent Power system restoration is an essential activity for grid resilience, where grid operators restart generators, re-establish transmission paths, and restore loads after a blackout event. With a goal of restoring electric service in the shortest time, the core decisions in restoration planning are to partition the grid into sub-networks, each of which has an initial power source for black-start (called sectionalization problem), and then restart all generators in each network (called generator startup sequencing problem or GSS) as soon as possible. Due to the complexity of each problem, the sectionalization and GSS problems are usually solved separately, often resulting in a sub-optimal solution. Our paper develops models and computational methods to solve the two problems simultaneously. We first study the computational complexity of the GSS problem and develop an efficient integer linear programming formulation. We then integrate the GSS problem with the sectionalization problem and develop an integer linear programming formulation for {the parallel power system restoration (PPSR) problem} to find exact optimal solutions. To solve larger systems, we then develop bounding approaches that find good upper and lower bounds efficiently. {Finally, to address 
{computational challenges for very large power grids}, we develop a randomized 
approach to find a high-quality feasible solution quickly.} Our computational experiments demonstrate that the proposed approaches are able to find good solutions for PPSR in up to 2000-bus systems. 
}


\maketitle

\section{Introduction}
In August 2003, 50 million people lost power for up to two days in the biggest blackout in North American history. The event contributed to at least 11 deaths and cost an estimated \$6 billion \citep{Conrad2006}.  After a devastating winter storm in February 2021, the recent Texas power crisis was even more deadly. More than 4.5 million homes and businesses were left without power, some for several days \citep{texas2021S}, and at least 151 people lost their lives directly or indirectly \citep{texas2021}. Damages from the blackouts were estimated at \$195 billion, making them the costliest disaster in Texas history \citep{texas2021F}. The enormous impact of an extended power system blackout highlights the need for a more resilient grid that can recover rapidly from disruptions due to extreme weather events or malicious attacks \citep{Adibi2006}. 

{Power system restoration aims to restore electricity service after a partial or complete blackout event. {Restoration includes transmission system restoration and distribution system restoration. This work focuses on transmission system restoration planning because transmission system restoration is both more crucial for electricity service recovery and more complex in that it involves coordination of multiple assets and stakeholders.} During restoration, system operators use initial power sources (called black-start generators, i.e., generators that can start by themselves without power from the grid, such as hydro generators and thermal generators with backup fuel and power) to start non-black-start generators (which need external power to start), energize the transmission network, and pick up loads. Restoration is a complicated process where operators must not only consider technical and safety requirements, but also demand close collaboration among utility companies and RTO/ISOs. Therefore, a carefully designed restoration plan is crucial for the success of restoration activities. 
} 

{A fundamental sub-problem in restoration planning is to sectionalize the network and restart generators.} 
{The sectionalization problem partitions}
a large system into a set of smaller subsystems, each of which has an initial power source. 
Sectionalizing the system is beneficial for the following reasons: (1) Black-start resources are scarce. Partitioning the system into subsystems can better utilize the black-start resources and achieve a faster restoration with subsystems restored in parallel. (2) Generators lose synchronism in a large scale blackout. 
{To avoid severe voltage fluctuations it is often practical to partition the generators into multiple smaller self-supporting subsystems or islands before re-synchronizing them.} 
{Within each subsystem or island, the initial power source (black-start generator), is used to energize transmission paths, crank (restart) non-black-start generators, and restore generation capability. While only blackstart generators provide initial power, non-blackstart generators can also provide power once they are online to assist in restarting other generators. Before they restart, however, each non-blackstart generator consumes certain amount of cranking power for a given duration of time to restart (see Section \ref{sect:gss} for details). This makes the resulting scheduling problem complicated.} 

The literature has several papers that use sectionalizing and mixed-integer programming to model the problem. In \cite{wang2011}, the system is sectionalized so that the load and generation are balanced. 
In \cite{WAMS}, the system is sectionalized so that each island is fully observable by using Wide Area Measurement System.
{\citet{lin2011} proposed a method to partition the systems for restoration based on the community structure of complex network theory. \citet{lin2015} proposed an optimization model for optimal sectionalization of restoration subsystems with a special emphasis on the coordination of generator ramping and the pickup of critical loads.} \cite{sun} formulated and solved a mixed-integer linear program to optimize the startup sequence of generators assuming a copper plate model for the grid. 
{\citet{zhang2014} determined the unit start-up sequence to maximize restored generation capacity. \citet{gu2012} treated the unit start-up as a two layer restoration process; network-layer unit restarting and plant-layer unit restarting. \citet{wang2009} proposed a multi-objective optimization model for the unit start-up sequence, system-partitioning strategy and
time requirements. To reduce the optimization scale, \citet{zhu2014} employed a two-phase method with a lower solution quality.} 
{
\cite{pascal} solved the power system restoration problem with a linear programming approximation of the AC power flow equations introduced by \cite{pascal2}. }

{However, the literature above avoided integrating the generator startup sequencing problem with the sectionalization problem, given that each of them is a highly complex combinatorial optimization problem. Solving the sectionalization and GSS problems separately can result in sub-optimal solutions because the sectionalization results, the set of generators in an island, directly impact the restarting scheduling, and thus the total restoration time. For example, allocating too many non-blackstart generators that demand significant cranking power allocated to one island can potentially prolong the restoration time for that island. }
\cite{restoration} developed the first mixed integer linear programming formulation of the PPSR problem over a continuous time horizon 
{integrating the restoration problem with the sectionalization problem.} 
{However, their model takes a long time to solve resulting in several large networks that remain unsolved.} 

The main contribution of our paper is to propose models that integrate the sectionalization and GSS into a single optimization problem and develop efficient bounding approaches that allow the solution of large instances that represent regional and interconnection-level grids.
{Besides the operational decisions considered in this paper, power system restoration planning involves many steps, such as plan validation to check voltage and frequency stability, dynamic security, and other power-system physics. We refer readers to \cite{restoration}, \cite{317561}, \cite{LINDENMEYER2001219}, \cite{8939530}, \cite{pascal2}, \cite{pascal1} and \cite{pascal} for other critical issues in power system restoration. Whereas these considerations are important for a restoration plan, our paper focuses on the PPSR problem. }

The rest of the paper is arranged as follows: 
In Section~\ref{sec:formulations}, we show that the GSS problem is NP-hard and develop a strong integer linear programming formulation. Integrating the GSS formulation with a sectionalizing formulation, we introduce an integer linear programming formulation of the PPSR problem. 
In Section~\ref{sec:enhancements}, we develop upper bounding and lower bounding approaches to speed up the integer linear programming formulation. The bounding approaches not only provide a good restoration plan, but also the maximum gap between the proposed plan and the optimal solution. In Section~\ref{sec:response}, we develop a probabilistic method, to solve large scale system with thousands of buses, e.g., at RTO/ISO level.  
Section~\ref{e:nemExp} presents numerical experiments for the exact integer programming approach, the upper and lower bounding approaches, and the probabilistic method. 
The experiments show that the proposed approaches are relatively fast and provide high quality solutions for restoration planing problems on targeted systems. 



\section{Integrated approach for power system restoration}\label{sec:formulations}
The power system is a network with nodes representing the buses and edges representing the transmission lines connecting the buses. 
{A bus can have generators, loads, or nothing as an intermediate bus.} The \emph{sectionalization problem} partitions the system into subsystems or {\em islands}, each with an initial power source.
In a sectionalizing plan, the initial power sources are black start (BS) generators, which can start on their own without power from the grid. The nodes with BS generators are called \emph{BS nodes} or \emph{root nodes}. Power from a BS generator is then used to provide cranking power to \emph{non-black-start} (NBS) generators that need external power to start. 

Each generator has its own time-varying generation capacity curve. An NBS generator needs cranking power to start, and the required cranking power at an NBS generator can be seen as a negative power capacity for that node until it starts generating power.
In every time period, the total capacity of a power system must be non-negative; i.e., all cranking power required by NBS generators in a time period must be supplied by other nodes in the network that are already generating power at that time. 
The \emph{generator startup sequencing} (GSS) problem starts with a single BS generator, satisfies the non-negative power capacity constraint over time, and sequences the start of generators to minimize the startup time of the last NBS generator, which we refer to as \emph{restoration time}.   
This paper introduces an integer linear programming formulation that integrates the GSS problem with sectionalizing plans\textemdash the parallel power system restoration or PPSR problem\textemdash the goal of which is to minimize the restoration time across all islands.
The PPSR problem starts NBS generators across all islands simultaneously with a goal of minimizing the NBS generator startup time across the entire system while satisfying the non-negative power capacity constraint {on each island} at all times.  

A practical PPSR problem may extend the set $I$ of NBS nodes to include \emph{critical loads} such as nuclear plants, traffic lights, hospitals, and police stations. Every node in $I$ must belong to some connected component having a BS node. 
However, the problem may relax the connectivity of certain nodes in an island. That is, some nodes may not be connected to BS nodes; they may be \emph{transshipment nodes} that connect other nodes to BS generators. 
The practical PPSR problem minimizes the startup time of cranking the last NBS generator or supplying the last critical load. 

In this section, we first develop an efficient integer programming formulation for the GSS problem and then integrate it into the sectionalization problem. 

\subsection{The generator startup sequencing (GSS)\label{sect:gss} problem}\label{sec:restoration}
We start by assuming a power system with a single BS generator. The generator startup sequencing problem minimizes the last startup time (restoration time) among NBS generators while supplying the critical loads and satisfying the non-negative total power capacity constraint in every time period.
{Note that each generator has its own ramping rate (the rate generator increases its power output) and maximal capacity.} Each generator $i$ has a cranking time of length $t_i^c$. We divide the time horizon into $T$ periods (1 period = 5 minutes in this paper). The start time of generator $i$ is represented by a binary variable $s_i(t)$ which takes the value 1 if generator $i$ starts in period $t$, 0 otherwise.   
During the cranking time, generator $i$ does not produce any power but consumes a constant cranking power $c_i$. In the capacity curve, the cranking power consumption is represented as a negative power output for generator $i$; i.e., $-c_i$. After the cranking period $t_i^c$, the generator starts producing power and ramps up for a time period $t_i^r$ to reach its maximum power output level $p_i$. The capacity curve for a generator during the ramping time is given as a non-negative increasing function. The capacity curve is thus a non-decreasing function on the time interval beginning from the start time $s_i$, and is determined by parameters $t_i^c,t_i^r,c_i,$ and $p_i$. 
{Note that, technically, a generator can change its power output to anywhere within its output limits. However, in order to minimize restoration time, we assume that each generator ramps up as fast as possible to reach its maximal capacity and then continues to generate at its maximal capacity. Therefore, once started, we assume that a generator follows its capacity curve to maximum and then stays at its maximal capacity. }   

\begin{figure}
\centering
\includegraphics[width=300pt]{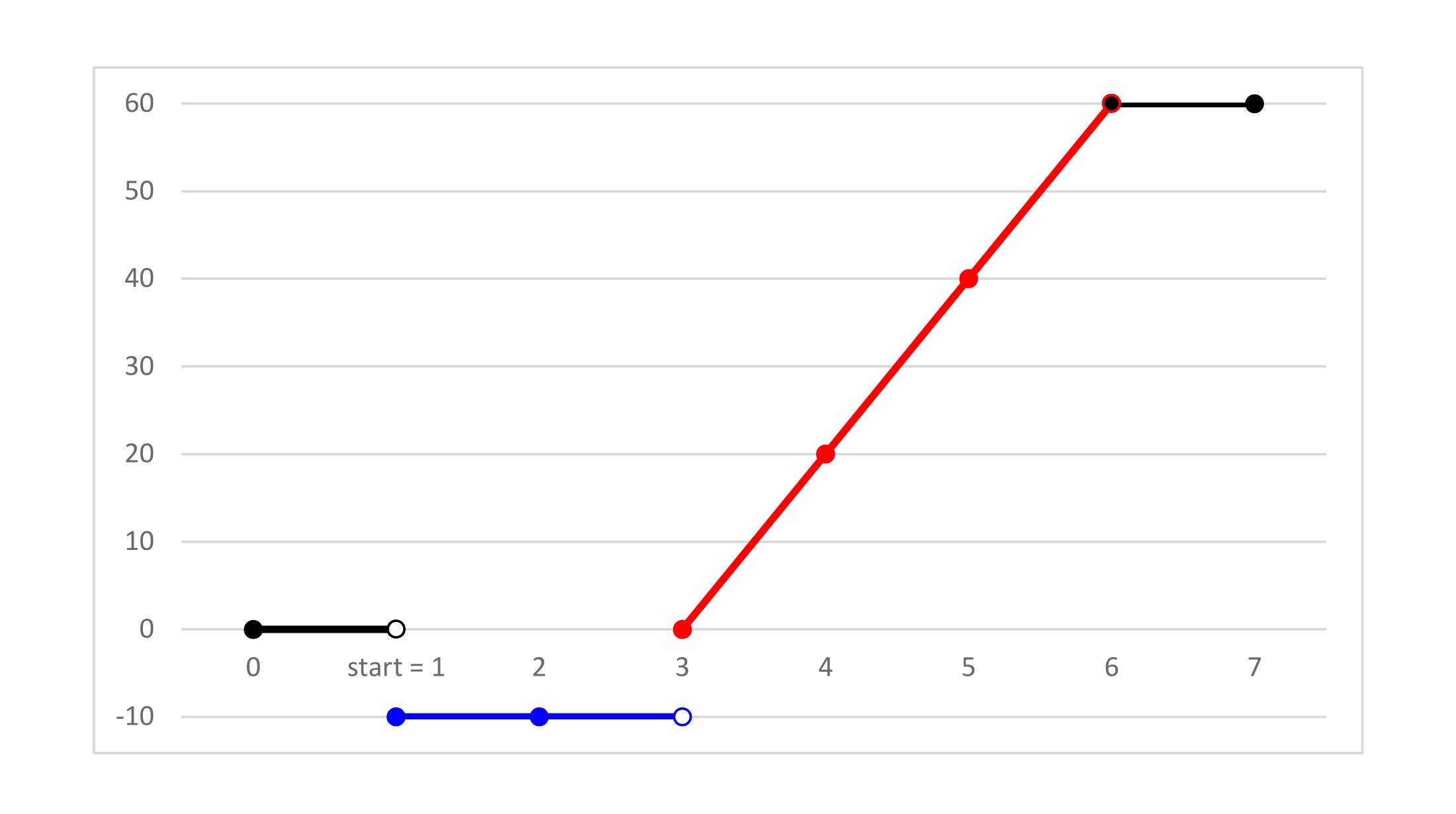}
\caption{The capacity curve $\boldsymbol{p_i(t)}$ of an NBS generator $\boldsymbol{i}$ with 10 MW of cranking power, two periods of cranking time (blue), three periods of ramping time (red), and a full capacity of 60 MW with the setting of $\boldsymbol{s_i (1) = 1}$. The cranking power needs to be supplied for two periods, and ramping starts at period 3.}\label{fig:cap}
\end{figure}

Figure~\ref{fig:cap} illustrates the capacity curve $p_i(t)$ of an NBS generator $i$ with 10 MW of cranking power, two periods of cranking time (blue), three periods of ramping time (red), and a full capacity of 60 MW. 
{
A critical load $l$ of demand $d$ is determined by the same set of parameters $t_l^c = T,t_l^r = 0,c_l = d, p_l = 0$. As a result, it behaves like an NBS node in our formulation of the PPSR problem. Except for case studies with critical loads, we assume that the set $I$ contains only NBS nodes.
}


Assuming one BS node with capacity curve $r(t)\geq 0$ in a connected power system, we first develop the generator startup sequencing (GSS) problem. The GSS problem aims to crank all NBS generators and supply all critical loads while keeping total capacity of the system non-negative over all time periods. The objective is to minimize the start time of the last NBS generator  $i\in I$.  

\begin{figure}
\centering
\includegraphics[width=450pt]{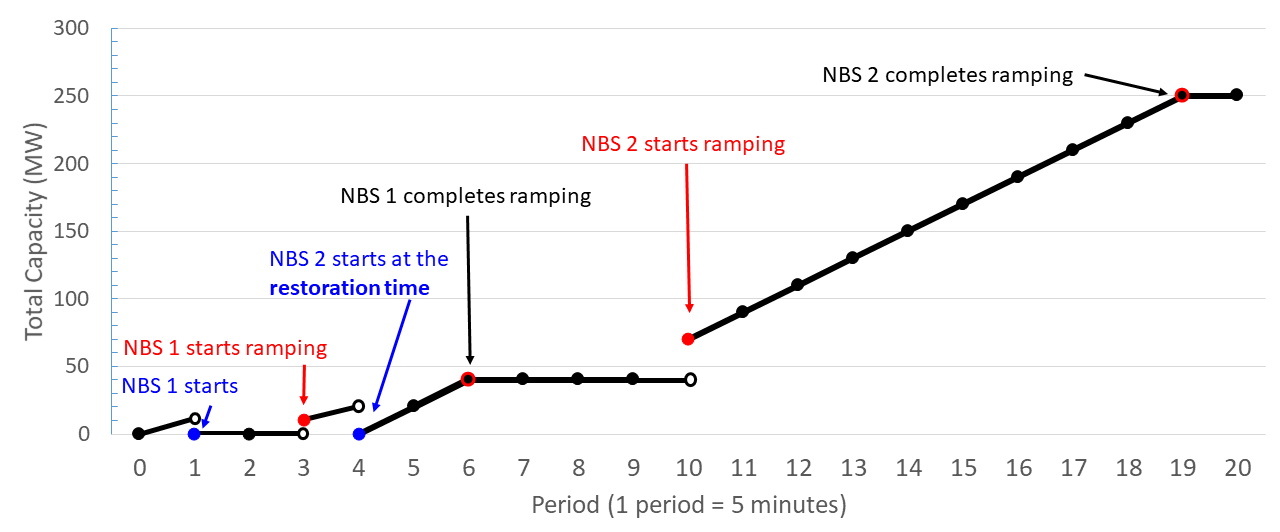}
\caption{Total capacity curve of the power system of BS, NBS 1 and NBS 2}\label{fig:totalCap}
\end{figure}

For example, consider a system with one BS generator and two NBS generators. Assume that the BS generator can immediately start ramping and produce its full capacity of 10 MW at period~1. More precisely, the parameters of the BS generator are $t^c = 0,t^r = 1,c = 0,p = 10$. One NBS generator (NBS~1) is assumed to have the capacity curve shown in Figure~\ref{fig:cap}. Let the cranking power of the other NBS generator (NBS~2) be 30 MW, and let its cranking and ramping times be six and nine periods respectively. In period 1, the BS power of 10 MW can not supply the 30 MW cranking power of NBS~2, but starts cranking NBS~1. In period~4, the generator NBS~1 produces 20 MW and is operating at capacity. The combined power of 20 MW from NBS~1 and 10 MW from BS equals 30 MW and can start cranking NBS~2. Thus, the minimum restoration time for this network is four periods. NBS~1 will complete ramping at period 6, and produce its full capacity of 60 MW while cranking NBS~2. At period 10, NBS~2 will start ramping after six periods of its cranking time. NBS~2 will produce its full capacity of 180 MW at period 19 after nine periods of its ramping time. The maximum total capacity of the power system will reach 250 MW (= 10 + 60 + 180) in period 19. See Figure~\ref{fig:totalCap} for the total capacity curve of this example.

We show that the generator startup sequencing problem is a reduction of the partition problem, and is thus NP-hard. The reduction shows that the feasibility of the generator startup sequencing problem is at least NP-complete.
\begin{theorem}
The generator startup sequencing (GSS) problem is NP-hard.
\end{theorem}

\paragraph{Proof.}
The partition problem aims to find a partition of integers $\{ \bar{p}_i :i\in I\}$ into subsets $S_1$ and $S_2$ where the sum of the values in each subset equals $p=\sum_{i\in I} p_i / 2$. Assume that the crank powers are $-p_i (1) = \bar{p}_i$ and all crank times are one period. Set $T=2$ and the starting power of the ramping time to 0. Then the partition problem trivially reduces to the generator startup sequencing problem. $\blacksquare$\\

Recall that $r(t)$ represents the capacity curve of the BS node. Define binary variables $s_i(t) = 1$ if the NBS generator $i$ is started at period $t$ and 0 otherwise. Let $RT$ be the restoration time, i.e., the start time of the last NBS generator. Given capacity curves $p_i(t)$ for each NBS generator $i$, we obtain an integer linear programming formulation of the generator startup sequencing problem as follows: 
\begin{eqnarray}
\min&&RT\label{eqn:objOneBS}\\
s.t.&&\sum_{t=1}^T s_{i} (t)=1\mbox{ for }i\in I\label{eqn:oneTime}\\
&&r(t) + \sum_{i\in I}\sum_{t^{-}=1}^t p_i (t - t^{-} + 1) \cdot s_{i} (t^{-}) \geq 0\mbox{ for }1\leq t\leq T\label{eqn:nonnegativeCapacity}\\
&&\sum_{t=1}^T t\cdot s_{i} (t) \leq RT\mbox{ for }i\in I\label{eqn:bottleneckOneBS}
\end{eqnarray}

Equation~\ref{eqn:oneTime} ensures that NBS generator $i$ starts once during the time horizon $T$. Equation~\ref{eqn:nonnegativeCapacity} ensures that the total capacity of the system is non-negative at all times. The left-hand side of Equation~\ref{eqn:bottleneckOneBS} is the start time of NBS generator $i$, and the formulation minimizes the restoration time $RT$ in Equation~\ref{eqn:objOneBS}.
Empirically, we find that this formulation  of the GSS problem is solved very quickly with a single BS generator. Throughout this paper, we use Equations~\ref{eqn:objOneBS}-\ref{eqn:bottleneckOneBS} to solve the GSS problem on each island or power system with a single BS generator.

\subsection{A Formulation for the Parallel Power System Restoration Problem}

Using Equations~\ref{e:binVarX}-\ref{eqn:degreeBound}, we develop an integer linear programming formulation of the parallel power system restoration (PPSR) problem that integrates the GSS problem and the sectionalization problem.
Let a power system after a blackout be represented by a connected graph $G$, where each node $v\in V$ represents a bus, and each edge $e\in E$ represents the transmission line between two buses. Let $J \in V$ be the set of BS nodes and let $I \in V$ be the set of NBS nodes. The nodes in $V \setminus (I \cup J)$ are transshipment nodes. 
Define binary variables as follows: 
\begin{equation}
x(v,j)\in\{ 0,1\}\mbox{ for }v\in V\mbox{ and }j\in J\label{e:binVarX}
\end{equation}
where $x(v,j) = 1$ implies that node $v$ belongs to the island of BS node $j\in J$ (or simply the island $j$). To ensure a sectionalizing plan in which each island has one BS node, and each NBS node belongs to one island, we include the following constraints:
\begin{eqnarray}
&&x(j,j)=1\mbox{ and }x(j,j')=0\mbox{ for }j\neq j'\in J\label{eqn:jInj}\\
&&\sum_{j\in J}x(v,j) = 1\mbox{ for }v\in I\label{eqn:partition}\\
&&\sum_{j\in J}x(v,j) \leq 1\mbox{ for }v\in V\setminus\left(I\cup J\right)\label{eqn:transNew}
\end{eqnarray}
Equation~\ref{eqn:jInj} stipulates that a BS node $j$ belongs to the island $j$. Equation~\ref{eqn:partition} stipulates that every NBS node  $v\in I$ belongs to exactly one island. 
{
Equation~\ref{eqn:transNew} stipulates that a transshipment node belongs to at most one island. Even in an island, a transshipment node does not have to be connected to a BS node, but may be used to connect NBS nodes to the BS node of the island.
}

Replacing $s_{i}(t)$ with 
\begin{equation}
s_{i}^j (t)\in\{0,1\}\mbox{ for }i\in I,j\in J,t\in\{1,2,...,T\} \label{e:startisland}   
\end{equation} 
we can extend the generator startup sequencing formulation to the PPSR problem in which $s_{i}^j (t) = 1$ indicates that generator $i$ in island $j$ starts at period $t$. An integer linear programming formulation of the generator startup sequencing component of the PPSR problem is the following:
\begin{eqnarray}
&&x(i,j) = \sum_{t=0}^T s_{i}^j (t)\mbox{ for }i\in I,j\in J\label{eqn:knapsack}\\
&&r^j (t) + \sum_{i\in I}\sum_{t^{-}\leq t} p_i (t - t^{-} + 1) \cdot s_{i}^j (t^{-}) \geq 0\mbox{ for }j\in J,t\in\{ 1,2,...,T\}\label{eqn:feasibleIsland}\\
&&\sum_{j\in J}\sum_{t=1}^T t\cdot s_{i}^j (t) \leq RT\mbox{ for }i\in I\label{eqn:balance}
\end{eqnarray}
The objective function is to minimize the restoration time $RT$:
\begin{equation}
\min RT\label{eqn:minBottelneck}
\end{equation}

{
To ensure connectivity of each island, we use a \emph{single commodity flow} model where all flow has to be sent to a single sink. A flow from a source $s$ to sink $t$ is a real valued function $f:V\times V\rightarrow \mathbb{R}^+$ that satisfies 
\begin{itemize}
    \item {\bf Capacity constraints:} For all $u,v\in V$, $f(u,v)\leq c(u,v)$
    \item {\bf Flow conservation:} For all $u\in V\setminus\{s,t\}$, $\sum_{v\in V} f(u,v)=0$
    \item {\bf Skew symmetry:} For all $u,v\in V$, $f(u,v)=-f(v,u)$
\end{itemize}
On each island, each NBS node (source) sends one unit of flow to the BS node (sink). 
The flow can only travel through the edges of the island. Thus, edge capacity $c(u,v)=0$ if $uv\not\in E$.
To reduce the number of flow variables, we assign an orientation $(u,v)$ to each edge $uv\in E$ such that each edge is assigned exactly one arbitrary orientation. We denote the set of oriented arcs (lines) as $L$. Then, the flow variable on a line may be positive or negative.  
}

{
For each $(u,v)\in L$ and $j \in J$, we define the binary variable $y^j(u,v)$ = 1 if line $(u,v)$ belongs to island $j$ (i.e., both ends $u,v$ belong to $j$) and 0 otherwise. In a sectionalizing plan, $y^j(u,v)=1$ for at most one $j\in J$ (i.e., a line can belong to at most one island). Indicating the network topology of the island, $y^j(u,v)$ are used to ensure that the flows go through the edges of the island.
Once an NBS node $i \in I$ is assigned to island $j$, there should be a path from $i$ to the BS node $j\in J$ which is ensured by one unit of flow sent from the NBS node to the BS node. The total flow into the BS node $j \in J$ must equal the number of NBS nodes $i\in I$ assigned to the island $j$. For each line $(u,v)\in L$, we define $|J|$ integer variables:
\begin{equation}
f^j (u,v)\in\mathbb{Z}\mbox{ for }j\in J\mbox{ and }(u,v)\in L \label{e:flowV-QL}
\end{equation}
each representing the number of units of flow on line $(u,v)$ flowing into BS node $j$.
A flow variable $f^j (u,v)$ for line $(u,v)$ may be a negative integer if the actual flow goes in the opposite direction  (from $v$ to $u$).
}

{
Using these variables, the sectionalization problem is formalized as follows:
\begin{eqnarray}
&&\sum_{(i,v)\in L}f^j (i,v) - \sum_{(u,i)\in L}f^j (u,i) = x(i,j)\mbox{ for }i\in I, j\in J\label{eqn:balanceAtI}\\
&&\sum_{(w,v)\in L}f^j (w,v) - \sum_{(u,w)\in L}f^j (u,w) = 0\mbox{ for }w\in V\setminus (I\cup J),j\in J\label{eqn:balanceAtTransEarly}\\
&&\sum_{(u,j)\in L}f^j (u,j) - \sum_{(j,v)\in L}f^j (j,v)= \sum_{i\in I} x(i,j)\mbox{ for }j\in J\label{eqn:balanceAtJ}\\
&&-y^j(u,v)|I|\leq f^j (u,v)\leq y^j (u,v)|I|\mbox{ for }(u,v)\in L,j\in J\label{eqn:flowBoundY}\\
&&\left.
\begin{array}{l}
y^j(u,v) \leq x(u,j)\\
y^j(u,v) \leq x(v,j)
\end{array}
\right\}
\mbox{ for }(u,v)\in L,j\in J\label{eqn:degreeBound}
\end{eqnarray}
}

{
Equation~\ref{eqn:balanceAtI} ensures that one unit of flow is sent from an NBS node $i \in I$ to the BS node $j \in J$ if and only if the NBS node $i$ is assigned to island $j$.
Equation~\ref{eqn:balanceAtTransEarly} stipulates flow conservation through a transshipment node $w\in V\setminus\left( I\cup J\right)$.
Equation~\ref{eqn:balanceAtJ} ensures that the total flow into a BS node $j \in J$ equals the number of the NBS nodes assigned to island $j$. 
Equation~\ref{eqn:flowBoundY} indicates that if $f^j (u,v) \neq 0$, then $y^j (u,v) = 1$, enforcing the following logic: If there is a flow on line $(u,v)\in L$ flowing through island $j$, then line $(u,v)\in L$ belongs to island $j$. 
Equation~\ref{eqn:degreeBound} indicates that if a line $(u,v)\in L$ belongs to island $j$, then both end nodes $u,v$ belong to island $j$. 
Equation~\ref{eqn:degreeBound} speeds up the sectionalization problem disaggregating the following constraint introduced by \citet{restoration}:
\begin{eqnarray*}
&&\sum_{(u,v)\in L}y^j(u,v) + \sum_{(v,w)\in L}y^j(v,w)\leq x(v,j) |\delta(v)|\mbox{ for }v\in V,j\in J,
\end{eqnarray*}
where $\delta (v)$ denotes the set of all the arcs connected to node $v$. 
}

Given a maximum time horizon $T$, the solution to the PPSR problem on a power network $G$ is expressed as follows: 
$$(x_{\mathrm{OPT}},s_{\mathrm{OPT}},RT_{\mathrm{OPT}})=\mathrm{PPSR} (G,T)$$
where $x_{\mathrm{OPT}}$, $s_{\mathrm{OPT}}$ and $RT_{\mathrm{OPT}}$ are the optimal sectionalizing plan, a schedule, and the optimal restoration time, respectively. 
In fact, $s_{\mathrm{OPT}}$ is optimal only on the \emph{bottleneck island} in which the restoration time is equal to the entire restoration time (there may be multiple bottleneck islands). Thus, $s_{\mathrm{OPT}}$ is frequently left out, and the solution expressed as 
$(x_{\mathrm{OPT}},RT_{\mathrm{OPT}})=\mathrm{PPSR} (G,T)$, $x_{\mathrm{OPT}}=\mathrm{PPSR} (G,T),$ or $RT_{\mathrm{OPT}}=\mathrm{PPSR} (G,T)$. 
Improving $s_{\mathrm{OPT}}$, a post-process solves the GSS problem for the optimal schedule on each island.
If time horizon $T$ does not have to be specified, $\mathrm{PPSR} (G,T)$ will be expressed as $\mathrm{PPSR} (G)$.

Our formulation can also incorporate other constraints that are important in practice.

\paragraph{Critical Time Interval:}
In starting steam units, it is important to coordinate certain critical time intervals, such as the maximum time interval beyond which certain thermal units cannot be safely restarted hot, or the minimum time interval required before a thermal unit can be started~\citep{Adibi1991}. Critical loads, such as off-site nuclear station power and critical gas infrastructures, must be served within required time limits. These critical time constraints can be modeled as follows:
\begin{eqnarray}
ET_i\leq \sum_{t=1}^T \sum_{j\in J} t \cdot s_{i}^j (t)\leq LT_i\label{e:cti}
\end{eqnarray}
where $ET_i$ and $LT_i$ are the earliest and latest start times for generator $i$, respectively.
We add such constraints to IEEE-30 bus system in Section~\ref{sec:caseStudy}.

\paragraph{Load-Generation Balancing:}
The load-generation balancing constraint requires that the difference between the total power generation and the total load within a section not exceed a certain amount $d$. Let $d_i$ denote the real power generation/load at bus $i$ ($d_i$ is positive for power generation and negative for load). The balancing constraint can be expressed as follows:
\begin{eqnarray}
-d\leq \sum_{i\in V\setminus J} d_i x(i,j) \leq d\mbox{ for }j\in J
\end{eqnarray}
In general, load-generation balancing is one of the major requirements for frequency stability, although this paper does not apply this constraint to any case study.

\section{Obtaining bounds for the PPSR formulation}\label{sec:enhancements}
Whereas the integer linear programming formulation (\ref{e:binVarX})-(\ref{eqn:degreeBound}) of the PPSR problem can be solved quickly for small networks, it can take a very long time for larger networks. To deal with larger networks, this section develops lower and upper bounding procedures for the integer linear programming formulation (\ref{e:binVarX})-(\ref{eqn:degreeBound}). The upper and lower bounds can be obtained significantly faster than solving the integer programming formulation to optimality. The bounds can be used to obtain high quality solutions for larger problems quickly.

{
For a minimization problem, a lower bound problem is constructed by dropping off some constraints from the original problem so that a feasible solution to the original problem is also a feasible solution of the lower bound problem, but not vice versa. In contrast, an upper bound problem is created by imposing more restrictions on the original problem so that a feasible solution for the upper bound problem is also feasible for the original problem.  
}

{More specifically, we construct the lower bound problem by aggregating all BS generators into a single node. This eliminates the need to partition the system in the lower bound problem. Since the lower bound problem is essentially a GSS problem with a single BS node, it can be solved efficiently. It is easy to verify that any feasible solution to the original PPSR problem is a feasible solution to the lower bound problem but not vice versa. A solution to the lower bound problem thus provides a lower bound for the original PPSR problem.}

{We construct the upper bound problem by restricting the power network to be a spanning tree of the original network. The resulting partition of a tree network is significantly easier to solve. We then run heuristics using local search to improve solution quality. It is not hard to see that a feasible solution to the upper bound problem is a feasible to the original PPSR problem and thus provides an upper bound for the original PPSR problem.}  

{Since the upper bound and lower bound problem can be solved independently, a parallel procedure can be devised to save computational time. }


\subsection{Obtaining lower bounds for the PPSR formulation}
The lower bounds are obtained using two observations:
\begin{itemize}
\item The integer linear programming formulation of the GSS problem with a single BS generator, given in Equations~\ref{eqn:objOneBS}-\ref{eqn:bottleneckOneBS}, can be solved quickly. This fact allows us to calculate a lower bound of the optimal value by solving the GSS problem with a (fictitious) central BS generator which aggregates all the BS power.  
\item Given a time horizon $T$, the infeasibility of $\mathrm{PPSR} (G,T)$ can be detected quickly when $T$ is shorter than the optimal restoration time $RT_{\mathrm{OPT}}$. If $\mathrm{PPSR} (G,T)$ is infeasible, we update the time horizon to $T+1$ and solve $\mathrm{PPSR} (G,T+1)$.
\end{itemize}

For an initial lower bound, we assume a single BS generator that aggregates the capacities of the BS generators at each time period $t\leq T$ and solve the GSS problem on the entire power system with the single central BS generator. We show that the optimal solution to the problem with a single central BS node is a lower bound for the original problem.
\begin{theorem}
The optimal value of the GSS problem on the entire power system with a single central BS generator (which aggregates the capacities of all BS generators) is a lower bound for the optimal value of the PPSR problem.
\end{theorem}
\paragraph{Proof:}
Assume the schedule $s_{\mathrm{OPT}}=(s_i^j:i\in I,j\in J)$ for the optimal solution $(x_{\mathrm{OPT}},s_{\mathrm{OPT}})$ of $\mathrm{PPSR} (G,T)$. Each $i\in I$ belongs to one island $j\in J$ where $s_i^j\neq 0$ has exactly one non-zero component $s_i^j (t) = 1$ at its start time $t$. For $i\in I$, define $s_i = s_i^{j(i)}$ if $i$ belongs to island $j(i)$. Because $s_i^j = 0$ (all components $s_i^j (t)$ are zero) if $i$ does \textit{not} belong to island $j$, the following is true:
$$s_i = s_i^{j(i)} = \sum_{j\in J} s_i^j$$
Let $r=\sum_{j\in J} r^j$ be the capacity curve of the central BS generator.
Then $(s_i:i\in I)$ is a feasible schedule for the GSS problem on an entire power system with a single central BS generator, as summing up Equation~\ref{eqn:feasibleIsland} over $j\in J$ implies Equation~\ref{eqn:nonnegativeCapacity}:
\begin{eqnarray*}
\lefteqn{r (t) + \sum_{i\in I}\sum_{t^{-}\leq t} p_i (t - t^{-} + 1) \cdot s_{i} (t^{-})}\\
&&= \sum_{j\in J} r^j (t) + \sum_{i\in I}\sum_{t^{-}\leq t} p_i (t - t^{-} + 1) \cdot \left(\sum_{j\in J} s_{i}^j (t^{-}) \right)\\
&&= \sum_{j\in J} r^j (t) + \sum_{i\in I}\sum_{t^{-}\leq t}\sum_{j\in J} p_i (t - t^{-} + 1) \cdot s_{i}^j (t^{-}) \\
&&= \sum_{j\in J} r^j (t) + \sum_{j\in J}\sum_{i\in I}\sum_{t^{-}\leq t} p_i (t - t^{-} + 1) \cdot s_{i}^j (t^{-}) \\
&&= \sum_{j\in J} \left( r^j (t) + \sum_{i\in I}\sum_{t^{-}\leq t} p_i (t - t^{-} + 1) \cdot s_{i}^j (t^{-})\right) \geq 0 \ \blacksquare
\end{eqnarray*}

\noindent
Our empirical results show that the lower bound obtained using this approach is fairly tight. We also use the initial lower bound to set an initial not-too-long time horizon (upper bound) for the (upper) bounding approach.

Our computational results show that the integer linear programming formulation $\mathrm{PPSR} (G,T)$ quickly detects the infeasibility of the problem instance if the time horizon $T$ is shorter than the optimal restoration time. For the lower bounding approach, we start with a lower bound of the optimal restoration time and increase $T$ by one unit each time we detect infeasibility. 

\begin{figure}
\centering
\includegraphics[width=15cm]{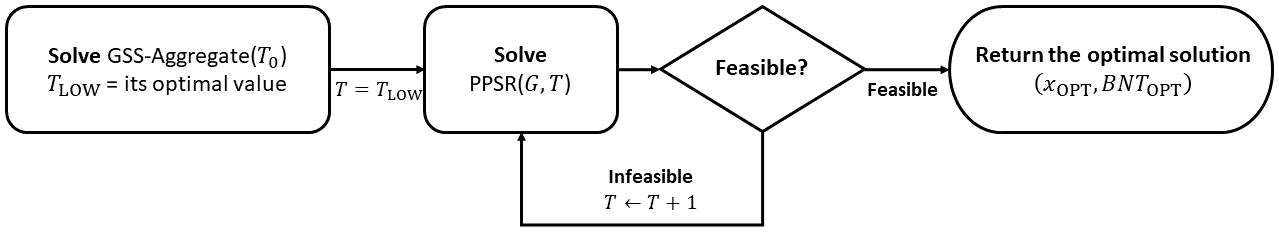}
\caption{The lower bounding approach}\label{fig:bottomup}
\end{figure}

In Figure~\ref{fig:bottomup}, we show the flowchart of the lower bounding approach. The first step aggregates total BS power into a single central BS generator and solves the GSS problem $\mbox{GSS-Aggregate} (T_0)$, where $T_0$ is an initial time horizon that is long enough. Let $T_{\mathrm{LOW}}$ denote the optimal value of $\mbox{GSS-Aggregate} (T_0)$. $T_{\mathrm{LOW}}$ represents a lower bound for the optimal solution for $\mathrm{PPSR}(G,T)$. Then solve $\mathrm{PPSR}(G,T),$ increasing the time horizon $T\geq T_{\mathrm{LOW}}$ one unit at a time until $\mathrm{PPSR}(G,T)$ is feasible. The first value of $T$ for which $\mathrm{PPSR}(G,T)$ is feasible equals the optimal restoration time. Let $(x_{\mathrm{OPT}},RT_{\mathrm{OPT}})$ be the optimal solution for $\mathrm{PPSR}(G,T)$.  

\subsection{Upper bounds}\label{s:upBound}
{
For the upper bounding approaches, we identify good feasible solutions beginning with the initial time horizon $T=2T_{LOW}$. Our empirical results show that $T=2T_{LOW}$ is typically an upper bound for the optimal solution.
} One iteration of the whole upper bounding approach consists of the following three steps:
\begin{enumerate}
\item Solving the PPSR problem on a random spanning tree: Identify a random spanning tree $RST_0$ of the entire system. Obtain an upper bound solution by solving the tree-partitioning formulation $\mathrm{PPSRT}\left(RST_0\right)$, which is defined by Equations \ref{e:binVarX}-\ref{eqn:minBottelneck} and \ref{eqn:integralTree}.  $\mathrm{PPSRT}\left(RST_0\right)$ can be solved very quickly on a tree because it has the minimum number of edges needed to maintain connectivity. The tree-partition provides a sectionalizing plan $x_0$ of the entire system.
\item Improving the solution using local search: Use local search to improve the upper bound solution. The local search procedure first merges two islands with the longest and the shortest restoration times into one island  {that induces a connected subgraph} with two BS nodes. The procedure then sectionalizes the merged island into two islands with the minimum restoration time on the merged island. {Repeat the process until there is no improvement on the longer of the restoration times for the two resulting islands.} Let $x_{LS}$ be the improved sectionalizing plan and $RT_{LS}$ the restoration time of the plan.
\item Improving the local search solution using a subgraph representing the best known plan: Construct a subgraph $G'$ which represents the sectionalizing plan $x_{\mathrm{LS}}$ given by the local search and solve $\mathrm{PPSR}(G',RT_{\mathrm{LS}})$. The islands of $x_{\mathrm{LS}}$ are represented by the breadth-first-search (BFS) trees of the islands. Along with the BFS trees, the cut edges of $x_{\mathrm{LS}}$ between the islands construct $G'$. Warm-start $\mathrm{PPSR}(G',RT_{\mathrm{LS}})$ with $x_{\mathrm{LS}}$. {The BFS tree is the shortest path tree of each island. Connecting every node to the BS node by the shortest path in the island, the BFS tree represents the connectivity structure of the island with the minimum number of edges.}
\end{enumerate}

Performing multiple iterations simultaneously starting with different random spanning trees $RST_0$ significantly speeds up the availability of a high-quality upper bound for the PPSR problem. 
{
It is easy to assign  transshipment nodes which are not connected to any BS nodes to islands so that the resulting islands induce connected subgraphs.
The sectionalization problem is thus equivalent to the \emph{centered network partition problem} to partition a given network into connected sub-networks, each with one center (i.e., one root node). In general, we may assume that $I=V\setminus J$ as the capacity of a transshipment node $v$ is zero (i.e., $t^c_v=t^r_v=c_v=p_v=0$).
}

We now discuss each of the three steps in our upper bounding approach in greater detail.

\subsubsection{Solving the PPSR problem on a random spanning tree.}\label{sec:treeHeu1}We use the integer linear programming formulation for the centered network partition problem on a tree developed by 
\cite{TreePartition} to quickly solve PPSR on a tree. \cite{TreePartition} showed that their formulation is integral.
\begin{lemma}[\citealt{TreePartition}]\label{eqn:treeForm}
Assume that $I=V\setminus J$. If $G$ is a tree, the convex hull of the integer solutions $x(i,j)$ is the polyhedron defined by Equations~\ref{eqn:jInj}-\ref{eqn:partition} along with non-negativity and the following constraints:
For $i\in V\setminus J$ and $j\in J$,
\begin{eqnarray}
x(i,j)\leq x(i',j)\mbox{ whenever $i'$ is the node adjacent to $i$ in the unique path from $i$ to $j$}.\label{eqn:integralTree}
\end{eqnarray}
\end{lemma}
{
Equation~\ref{eqn:integralTree} ensures that the intermediate nodes $i'$ between $i\in I$ and $j\in J$ belong to island $j$ if $i$ belongs to the island $j$. 
It is well-defined because there is exactly one path between two nodes of a tree.
}

{
Along with Equations~\ref{e:binVarX}-\ref{eqn:minBottelneck} the tree partitioning formulation of Equation~\ref{eqn:integralTree} is denoted by $\mathrm{PPSRT}$. The first upper bounding step solves $\mathrm{PPSRT} (RST_0,T_0=2T_{\mathrm{LOW}})$ on a random spanning tree $RST_0$ of the entire power system. The random spanning tree $RST_0$ can be found by assigning a random weight $w(u,v)\in U[0,1]$ to each edge $(u,v)\in E$ and solving the minimum weight spanning tree problem.
}
We use time horizon $T_0=2T_{\mathrm{LOW}}$ because our empirical results show that it is typically an upper bound of the optimal value of the PPSR problem for the entire power system. Let $x_0$ be the sectionalizing plan provided by this solution. Since $x_0$ is a partition of the spanning tree $RST_0$ into connected sub-trees, every island induces a connected subgraph of the power system.

\subsubsection{Improving the solution using local search.}\label{sec:LS}
{
Given a feasible sectionalizing plan $x_0$, the local search method picks a pair of islands which together induce a connected subgraph and solves the PPSR problem on the connected subgraph with two BS generators. The pair of islands chosen are those with the longest restoration time and a connected island with a shorter restoration time. The goal of local search is to move some NBS nodes from the island with long restoration time to the island with short restoration time, thus reducing the longest restoration time. 
}

\begin{algorithm}
\caption{The local search method (LS)}
\label{alg:LS}
\hspace*{\algorithmicindent} \textbf{Input:} A feasible sectionalizing plan and its restoration time $\left( x_0, RT_0\right)$ to $\mathrm{PPSR}(G)$\\
\hspace*{\algorithmicindent} \textbf{Output:} 
A feasible sectionalizing plan and its restoration time 
$\left( x_{\mathrm{LS}},RT_{\mathrm{LS}}\right)=\mathrm{LS} \left( x_0,RT_0\right)$ with $RT_{\mathrm{LS}}\leq RT_0$

\begin{algorithmic}[1]
\STATE (Initialization) Set $x_{\mathrm{Temp}}=x_0$ and solve $RT_{\mathrm{Temp}}^j=\mathrm{GSS}\left( V_j\right)$ for $j\in J$; set {\sf improved} = {\sf Yes}
\label{Line:Initialize}
\WHILE{{\sf improved} = {\sf Yes}}
\STATE {\sf improved} = {\sf No}; $j_{\max}=\argmax_{j\in J}\left\{ RT_{\mathrm{Temp}}^{j}\right\}$; $RT_{\max}=\max_{j\in J}\left\{ RT_{\mathrm{Temp}}^j\right\}$ \label{Line:fixMax}
\FORALL{$j_{\mathrm{small}}\in J\setminus \left\{ j_{\max}\right\}$ in the increasing order of $RT_{\mathrm{Temp}}^j$}\label{line:small}
\IF{$G\left[ V_{\max}\cup V_{\mathrm{small}}\right]$ is connected}
\STATE solve $\left( x_{\mathrm{TWO}},RT_{\mathrm{TWO}}\right)=\mathrm{PPSR}\left( G\left[ V_{\max}\cup V_{\mathrm{small}}\right],RT_{\max}\right)$

\IF{$RT_{\mathrm{TWO}}<RT_{\max}$}
\STATE {\sf improved} = {\sf Yes}
\STATE $x_{\mathrm{TWO}}$ updates two islands $V_{\max},V_{\mathrm{small}}$ and $x_{\mathrm{Temp}}$ on the two islands
\STATE solve and update $RT_{\mathrm{Temp}}^j = \mathrm{GSS} \left( V_j\right)$ for $j\in\left\{ j_{\max},j_{\mathrm{small}}\right\}$
\STATE break this {\bf forall}-loop
\ENDIF
\ENDIF
\ENDFOR\label{Line:EOL}
\ENDWHILE
\STATE return $\left( x_{\mathrm{LS}},RT_{\mathrm{LS}}\right)=\left( x_{\mathrm{Temp}},RT_{\max}\right)$
\end{algorithmic}
\end{algorithm}

Algorithm~\ref{alg:LS} provides the details of our local search method.
Given a feasible sectionalizing plan $x_0$, the local search method first sets the initial plan $x_0=x_{\mathrm{Temp}}$ and solves the GSS problem $RT_{\mathrm{Temp}}^j=\mathrm{GSS}(V_j)$ on the nodes $V_j$ of each island $j\in J$ (Line~\ref{Line:Initialize}).
The algorithm then repeats the following steps:
Fix the island $j_{\max}$ that defines the overall restoration time (Line~\ref{Line:fixMax}).  Enumerate the other islands in increasing order of their own restoration times. {For each island that induces a connected subgraph with the fixed island, solve the PPSR problem for the connected subgraph induced by the two islands} (Line~\ref{line:small}-Line~\ref{Line:EOL}). If the restoration time decreases ({\sf improved = yes}), repeat the process (fixing the bottleneck island and enumerating the other islands). If the restoration time does not improve ({\sf improved = No}), stop.


Given a feasible sectionalizing plan $x_0$ and its restoration time $RT_0$, the local search method provides a potentially better solution denoted by the following:
$$(x_{\mathrm{LS}},RT_{\mathrm{LS}})=\mathrm{LS}(x_0,RT_0)$$

\subsubsection{Improving the local search solution using a subgraph representing the best known plan.}
Denote by $B^j_{\mathrm{LS}}$ the breadth-first-search (BFS) tree (rooted at the BS node $j$) of the subgraph induced by every island $j\in J$ for the best known plan $\left( x_{\mathrm{LS}},RT_{\mathrm{LS}}\right)$.
Adding the cut edges $C\left(x_{\mathrm{LS}}\right)$ to the tree edges $\bigcup_{j\in J}E\left(B^j_{\mathrm{LS}}\right)$, we solve $\mathrm{PPSR} (G',RT_{\mathrm{LS}})$ on the {connected} sub-graph $G' = (V, E')$ with $E' = \left(\bigcup_{j\in J}E\left(B^j_{\mathrm{LS}}\right)\right) \cup C\left(x_{\mathrm{LS}}\right)$. 
{
The subgraph $G'$ connects the BFS trees within the islands using the cut edges across the islands.
}
Given the small number of edges in $|E\left( G'\right)|$, the PPSR problem on $G'$ is solved faster than on the entire power system $G$.
A warm start with an initial incumbent solution $x_{\mathrm{LS}}$ speeds up $\mathrm{PPSR} (G',RT_{\mathrm{LS}})$ even more. 
This heuristic frequently results in the upper bounding approach giving the exact optimality of the entire power system $G$. The resulting restoration time $RT'$ is at most $RT_{\mathrm{LS}}$ as shown next.

\begin{theorem}\label{thm:BFST}
Assume that $I = V\setminus J$. 
Let $x_{\mathrm{LS}}$ be a feasible sectionalizing plan for $\mathrm{PPSR} (G)$, and let $RT_{\mathrm{LS}}$ be the restoration time of the plan. Then $$RT'=\mathrm{PPSR} (G')\leq RT_{\mathrm{LS}}$$
where $V(G')=V(G)$ and $E(G') = \left(\bigcup_{j\in J} E(B^j_{\mathrm{LS}})\right) \cup C(x_{\mathrm{LS}})$.
\end{theorem}

\paragraph{Proof:}
The breadth-first-search trees $\left\{ B_{\mathrm{LS}}^j : j\in J\right\}$ are sub-trees of $G'$, and $x_{\mathrm{LS}}$ is also a feasible sectionalizing plan for $\mathrm{PPSR}(G')$.
$\blacksquare$\\


\section{{A randomized approach for large-scale instances}}\label{sec:response}
The bounding approach detailed in the previous section allows us to solve problems with up to a few hundred nodes in a reasonable amount of time. However, the bounding approach takes too long for ultra-large power systems (systems with 1000 or more buses). The PPSR problem takes a long time even on a spanning tree of a thousand nodes (e.g., $\mathrm{PPSRT}(RST_0)$ takes over 15 hours of computational time on PEG-1354). As a result, even after an hour, the upper bounding approach is unable to start with an initial sectionalizing plan found on a random spanning tree. 
\begin{algorithm}
\caption{Random Sectionalizing Plan}
\label{a:arr}
\hspace*{\algorithmicindent} \textbf{Input:} Power system $G$ with BS generators $J$ \\
\hspace*{\algorithmicindent} \textbf{Output:} Sectionalizing plan $x^{\mathrm{RSP}}$
\begin{algorithmic}[1]
\STATE (Initialization of sets) $V_j = \{ j\}$ for $j\in J$; $W=V\setminus J$ \label{l:initialize}
\STATE (Initialization of vector) $x^{\mathrm{RSP}} (i,j) = 1 - x^{\mathrm{RSP}} (j,j) = 0$ with $i\neq j$ for $i\in V$ and $j\in J$ \label{l:initialize2}
\WHILE{$W\neq \emptyset$}\label{l:whileStarts}
\STATE $C=\emptyset$
\FOR{$\left(w,j\right)\in W\times J$}\label{l:forLoop}
\IF{$V_j$ has a neighbor of $w$}
\STATE $C\leftarrow C\cup\{ (w,j)\}$
\STATE Observe $\tilde{x}(w,j) \in U\left[ 0,1\right]$
\ENDIF
\ENDFOR
\STATE $(w^{\max},j^{\max})=\argmax\left\{\tilde{x}(w,j):(w,j)\in C\right\}$
\STATE $x^{\mathrm{RSP}}\left( w^{\max}, j^{\max}\right) = 1$\label{l:observe}
\STATE $V_{j^{\max}} \leftarrow V_{j^{\max}}\cup\{ w^{max}\}$; $W \leftarrow W\setminus \{ w^{max}\}$
\ENDWHILE\label{l:whileEnds}
\RETURN $x^{\mathrm{RSP}}$ indicating connected partition $\left( V_j:j\in J\right)$
\end{algorithmic}
\end{algorithm}

For very large problems, a realistic goal is to find a high-quality solution quickly. This section develops a probabilistic method to find an initial feasible sectionalizing plan on a large scale power system. Local search discussed in Section~\ref{sec:LS} is then used to improve solution quality. This approach provides high quality solutions for large instances in a reasonable amount of time. 

A sectionalizing plan is a centered network partition; i.e., each island has one center (BS node) and induces a connected subgraph. A sectionalizing plan is \emph{feasible} if it provides a feasible solution to the PPSR problem. A sectionalizing plan may be infeasible, if the restoration time of an island exceeds the time horizon, or if a large NBS generator cannot be started by the other generators of the island. Our probabilistic method finds random sectionalizing plans and the local search method introduced in Section~\ref{sec:LS} improves the feasible ones among the sectionalizing plans found. The approach simultaneously performs multiple iterations, each of which first finds a random sectionalizing plan, which if feasible, is improved using local search.

\begin{algorithm}
\caption{Restoration Time of Sectionalizing Plan}
\label{a:evaluator}
\hspace*{\algorithmicindent}\textbf{Input:} Sectionalizing plan $x^{\mathrm{RSP}}$; BS nodes $J$; time horizon $T$\\
\hspace*{\algorithmicindent}\textbf{Output:} Restoration time $RT\leq T$ or $RT=\infty$ at island $BNI\in J$
\begin{algorithmic}[1]
\STATE (Initialization) $RT = 0$; $BNI = \emptyset$ 
\FOR{$j\in J$}
\STATE Solve the GSS problem on $V_j=\{ v\in V:x^{\mathrm{INT}}(v,j)=1\}$ with time horizon $T$ 
\STATE $RT_j\leq T$ if the GSS problem is feasible; $RT_j=\infty$ otherwise
\IF{$RT < RT_j$}
\STATE $RT\leftarrow RT_j$; $BNI\leftarrow j$
\ENDIF
\IF{$RT_j=\infty$}
\STATE {\bf break}
\ENDIF
\ENDFOR
\STATE {\bf return} restoration time $RT$ with bottleneck island $BNI\in J$
\end{algorithmic}
\end{algorithm}

Given a power system $G$ with the BS generators $J$, Algorithm~\ref{a:arr} provides a random sectionalizing plan $x^{\mathrm{RSP}}\in\{0,1\}^{V\times J}$.
In the algorithm, Lines~\ref{l:initialize} and \ref{l:initialize2} initialize the islands $V_j=\{ j\}$ of BS nodes $j\in J$, the nodes $W=V\setminus J$ which will be added to islands, and $x^{\mathrm{RSP}}$ which initially indicates the singleton connected partitions (i.e., $V_j = \{ j\}$ for $j\in J$) but will finally indicate a sectionalizing plan $\left( V_j:j\in J\right)$. Each node island combination $(w,j) \in W \times J$ is assigned a random weight $\tilde{x}(w,j)\in U\left[0,1\right]$, where $U\left[0,1\right]$ is the independent uniform distribution on the unit interval $[0,1]$. Each island $V_j, j \in J$ induces a connected subgraph throughout the algorithm. Among all nodes not yet assigned to islands (i.e., as long as $W \neq \emptyset$), Lines~\ref{l:whileStarts}-\ref{l:whileEnds} find a node $w^{\max}$ to be added to island $V_{j^{\max}}$ with the highest random value $\tilde{x}(w,j)$ across $\left(w,j\right)\in W\times J$ with $w^{max}$ having a neighbor in $V_j$. The node $w^{max}$ is then removed from the set $W$ and steps 3-14 are repeated until $W$ is empty. 

Once a random sectionalizing plan $x^{\mathrm{RSP}}$ is found by Algorithm~\ref{a:arr}, we can check infeasibility or obtain the restoration time of the sectionalizing plan using Algorithm~\ref{a:evaluator} within a reasonable time. We evaluate the restoration time $RT_j$ of each island $V_j$ by solving the GSS problem on the island. The integer linear programming formulation in Equations~\ref{eqn:objOneBS}-\ref{eqn:bottleneckOneBS} proves infeasibility or solves the GSS problem within a reasonable time although the GSS problem is NP-hard. If the GSS on all islands is feasible, the overall restoration time is $RT = RT \left( x^{\mathrm{RSP}}\right) = \max_{j\in J} RT_j$ at bottleneck island $BNI=BNI\left( x^{\mathrm{INT}}\right)$.

\begin{figure}
\centering
\includegraphics[width=15cm]{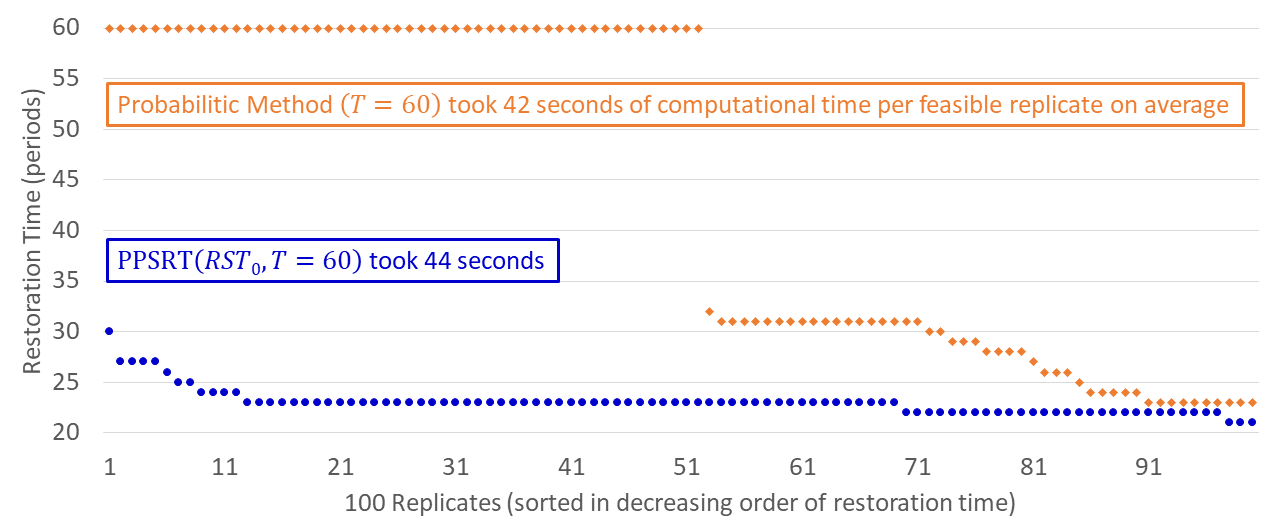}
\caption{Separating and integrating the sectionalization problem and the GSS problem on IEEE-118.}\label{fig:integrating}
\end{figure}
In practice, the sectionalizing plan found by Algorithm~\ref{a:arr} is often infeasible. For example, on 100 random sectionalizing plans (i.e., connected network partitions) found by Algorithm~\ref{a:arr} on IEEE-118 with a 60 period time horizon (see Figure~\ref{fig:integrating}), we find that more than 50 plans are infeasible in that either the restoration time exceeds the long time horizon, or there is a large NBS generator which cannot be started by the other generators of the island.


We perform multiple trials of Algorithms~\ref{a:arr} and \ref{a:evaluator} until a feasible sectionalizing plan is found. The first feasible sectionalizing plan found is then improved using the local search method described earlier in Section 3. We perform multiple iterations of this approach and select the best available solution.

\section{Numerical experiments}\label{e:nemExp}
In this section, we perform numerical experiments using the approaches introduced in Sections~\ref{sec:formulations}, \ref{sec:enhancements}, and \ref{sec:response}.
Our approaches are implemented in Python.
The integer linear programming formulations are solved using GUROBI solver version 9.0. 
The computational times reported in this section are CPU time plus the administration time (e.g., time to feed input data).
We used the Pittsburgh Supercomputer \citep{towns2014xsede} to perform parallel computing with eight threads and shared memory to solve each run of a problem.

{We select test systems of various sizes for our computational experiments (Table~\ref{tab:allData}). For example, SC-500 is a synthetic, regional scale system representing the power grid in South Carolina. PEG-2383 is a large scale system that represents 400, 220 and 110 kV networks of Polish power system, a regional transmission organization (RTO) level. As the top-level restoration planning is often performed at RTO,  the large scale test systems in this paper represent the most computationally challenging restoration planning scenarios.}

\subsection{Integer programming approach on small to medium scale power systems}\label{sec:caseStudy}
In this section, we solve power systems IEEE-30, IEEE-118, IEEE-300 and SC-500. We find that for small problems (IEEE-30 and IEEE-118), our IP formulation can directly be solved in a reasonable amount of time to obtain the optimal solution. For IEEE-118, using our bounding approach reduces the time needed to obtain the optimal solution. For larger problems (IEEE-300 and SC-500), the IP formulation takes too long to solve directly. The bounding approaches, however, provide high quality solutions to both problems in a reasonable amount of time. The details for each problem are provided next.

\begin{table}
\TABLE
{The power systems tested in this paper\label{tab:allData}}{
\centering
\scalebox{1}{
\begin{tabular}{|ccr|c|}
\hline
\up\down &ID &$|E|$ & Scale \\
\hline
\up\down &IEEE-30& 41 &  Small Scale\\
\up\down &IEEE-118&186 &   \\
\hdashline
\up\down &IEEE-300& 411 & Medium Scale \\
\up\down &SC-500&597 &   \\
\hline
\end{tabular}

\quad

\begin{tabular}{|ccr|c|}
\hline
\up\down &ID &$|E|$ & Scale \\
\hline
\up\down &PEG-1354&1,991 &  \\
\up\down &RTE-1888&2,531 & Large Scale\\
\up\down &RTE-1951&2,596 & \\
\up\down &PEG-2383&2,896 & \\
\hline
\end{tabular}

}
}
{Note: The number in the ID is the number of buses (nodes).}
\end{table}\label{tab:instances}

\subsubsection{Case study: IEEE-30.}
The test used a modified IEEE-30 bus system. To make this case computationally challenging, we added the same 25 generators, four of which (at buses 2, 11, 13, 24) are BS units, and the same four critical loads used in \cite{restoration}. The ranges of the parameters of the BS generators are the same as in \cite{restoration}. We also included additional constraints (critical time interval constraints) as in their paper. We created the non-black-start generator and the critical load parameter ranges in  proportion to the real demand at the buses. 
Whereas \cite{restoration} solved the network with piecewise linear capacity curve on a continuous time horizon in 503 seconds, our model solved a discrete time version of the integer linear programming formulation (Equations \ref{e:binVarX}-\ref{eqn:degreeBound} along with critical time interval constraint (Equation \ref{e:cti})) within 2 seconds. 
Given that we directly solved the integer formulation, we did not need to use any bounding approach to solve IEEE-30.

\subsubsection{Case study: IEEE-118.}\label{sec:IEEE-118}
The IEEE-118 bus system consists of 118 buses, 186 branches and 54 generators. We used the data from \cite{data118} for full capacities, cranking powers, ramping times and cranking times of the generating units, and critical loads (Class I load). For more details on the generator parameters, see Appendix. According to the optimal black start allocation solved by \cite{oren}, we assumed six BS nodes (buses 21, 22, 25, 28, 45 and 51). 
Despite using a fine time unit (1 period = 5 minutes)  (\cite{oren} used a coarse time unit of 1 period = 15 minutes), we obtained an optimal solution using our bounding approach in 302 seconds. {The importance of our enhancements is highlighted by the fact that solving the ILP formulation $\mathrm{PPSR}(G,T=38)$ itself without any enhancements takes 1,945 seconds.}

\begin{figure}
\centering
\includegraphics[width=15cm]{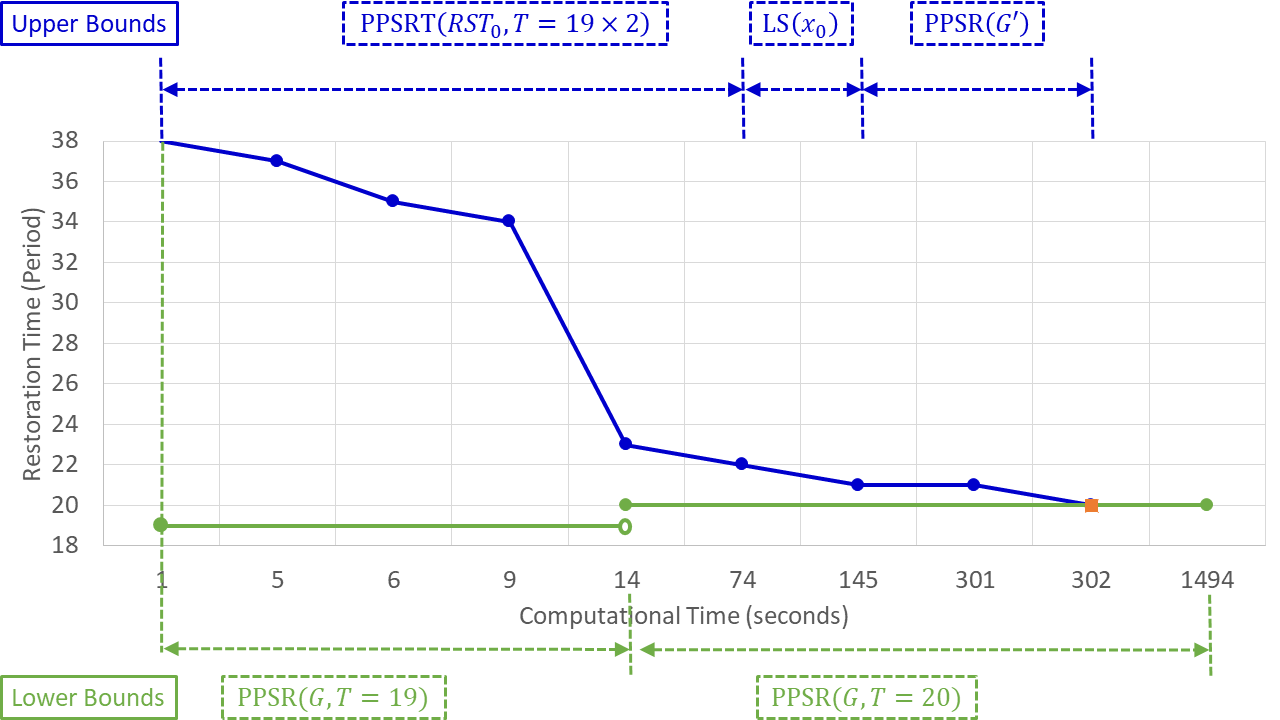}
\caption{Upper and lower bounds of the integer programming approach for IEEE-118. The problem is solved to exact optimality in 302 seconds.  \cite{restoration} took 5 hours 36 minutes to solve the same problem.}\label{fig:118}
\end{figure}

Figure~\ref{fig:118} shows the evolution of our approach for solving the IEEE-118 bus system to optimality. We performed the bounding approach introduced in Section~\ref{sec:enhancements}.
We first solved GSS-Aggregate$(T=60)$ obtaining a lower bound $T_{\mathrm{LOW}}=19$ in one second. Subsequently, the lower bounding approach showed $\mathrm{PPSR}(G,T=19)$ to be infeasible in 13 seconds, thus providing a lower bound of 20.
{The lower bounding approach itself took 14 seconds (= 1 + 13).}
Starting with time horizon $T_0=2T_{\mathrm{LOW}}=38$, the upper bounding approach solved $\mathrm{PPSRT}(RST_0)$, $\mathrm{LS}(x_0)$ and $\mathrm{PPSR}(G')$ finding the optimal solution $x_{\mathrm{OPT}}$ 
{in 302 seconds (= 1 + 73 + 71 + 157).} 
\cite{restoration} took  5 hours 36 minutes to solve IEEE-118 on a continuous time horizon. 

\subsubsection{Case study: IEEE-300 and SC-500.}\label{sec:simulation}
To check the limits of our bounding approach, we tested its performance on medium scale power systems IEEE-300 and South Carolina(SC)-500 from \citet{Matpower}. Since only limited information on parameters such as ramping rate and load over a certain time interval is given in that work, we used a simple substation placement and used IEEE-118 buses to generate problem instances as well as  crank power, crank time, full capacity and critical load on IEEE-300 and SC-500. The largest (smallest resp.) IEEE-118 generator was placed at the generator node of the largest (smallest resp.) parameter on the medium scale power systems.  

\begin{figure}
        \centering
        \begin{subfigure}[b]{0.475\textwidth}
            \centering
            \includegraphics[width=\textwidth]{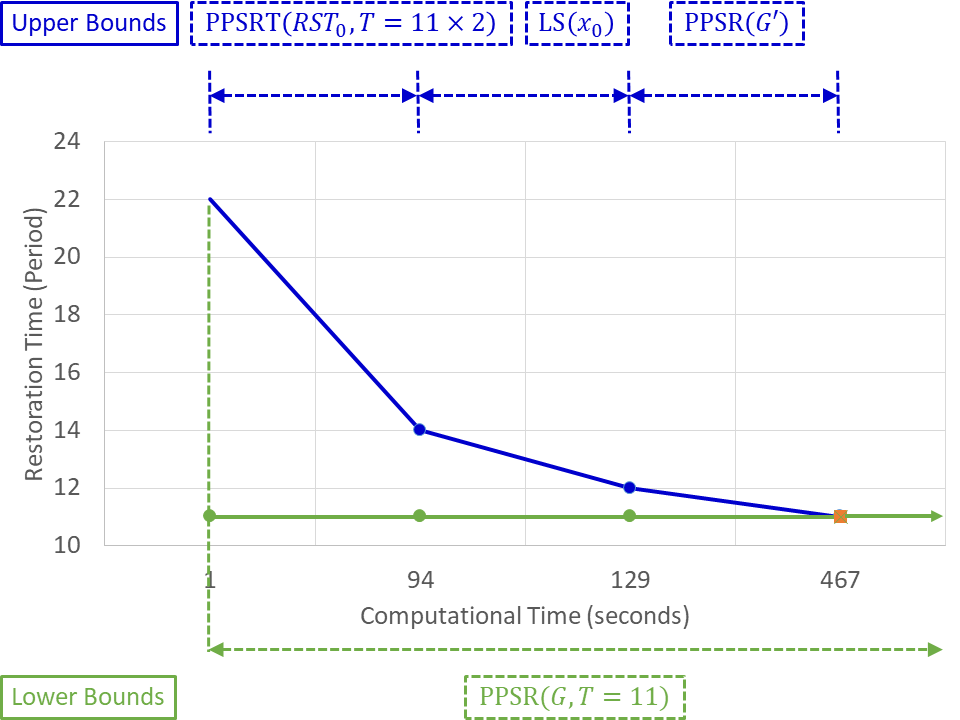}
            \caption{IEEE-300}    
            \label{IEEE-300}
        \end{subfigure}
        \hfill
        \begin{subfigure}[b]{0.475\textwidth}  
            \centering 
            \includegraphics[width=\textwidth]{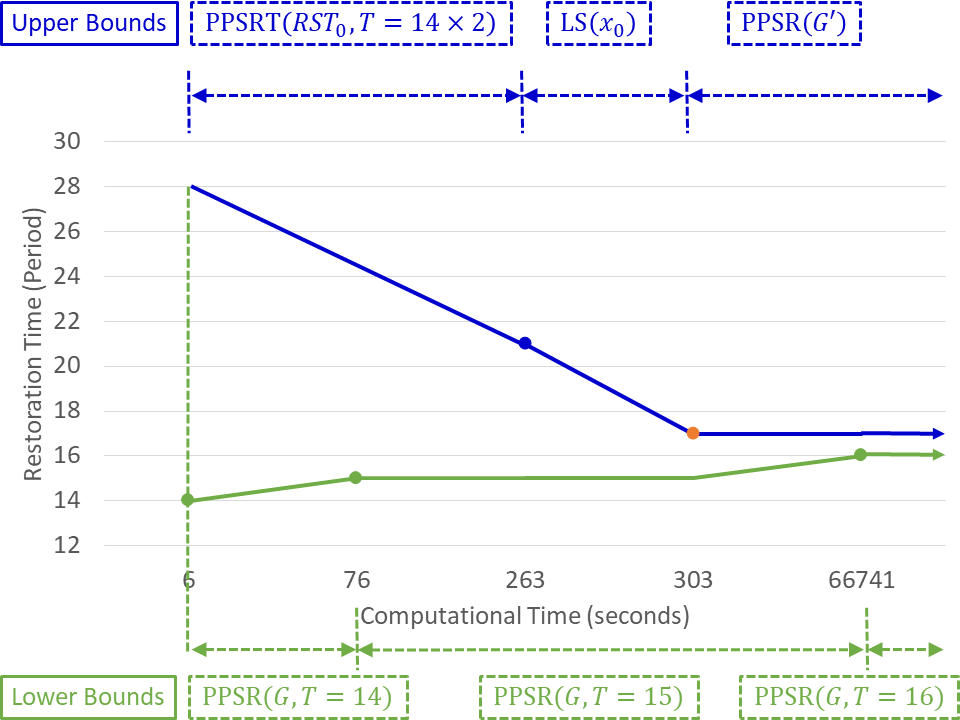}
            \caption{SC-500}    
            \label{SC-500}
        \end{subfigure}
        \caption{Integer Programming Approaches on Medium Scale Power Systems}\label{fig:largeScale}
\end{figure}

Since the ramping rates are only known for the generators of IEEE-300, and $P_{\max}$ is only known for the generators of SC-500, the generators of IEEE-118 are assigned to the buses of IEEE-300 so that their full capacities are proportional to their ramping rates (and proportional to $P_{\max}$ for the generators of SC-500). In a similar manner, we scaled up the critical loads of IEEE-118 to the loads of the medium scale power systems, proportional to their maximum loads over a certain time interval on IEEE-300 and proportional to $P_d$ on SC-500. The nodes with the highest degrees were chosen to be the BS nodes. For the feasibility of the instances, the BS capacities are set to the same  maximum BS capacity, 48.49 MW, equal to the largest BS of IEEE-118.

The results for both problems are provided in Figure~\ref{fig:largeScale}. 
On IEEE-300, the lower bounding approach using an aggregate BS node gave a lower bound of 11 periods very quickly. {In contrast, the lower bounding approach without aggregation solves the problem $\mathrm{PPSR}(G,T=11)$ in 24,721 seconds. This highlights the value of using aggregation for very large problems.}  {Within 467 seconds}, the upper bounding approach obtained the upper bound of 11 periods, which equals the lower bound and is thus optimal. On SC-500, we obtained a lower bound of 14 periods by solving GSS-Aggregate very quickly. Using a random spanning tree, the initial upper bound obtained was 21. The local search method quickly improved the upper bound by identifying a solution $x_{\mathrm{LS}}$ with a restoration time of $17$ periods. 
After 66,741 seconds, the lower bounding approach increased the lower bound to be at least 16. Thus, the upper bound of 17 periods has an optimality gap is at most 1 (= 17 $-$ 16) and is near optimal. We achieved the near optimal solution $x_{\mathrm{LS}}$ on SC-500 within a total of 303 seconds (= 6 + 257 + 40).

{
We then attempted an even larger problem, PEG-1354. In this case, solving $\mathrm{PPSRT}(RST_0)$ on a random spanning tree $RST_0$ itself took 54,319 seconds, showing the challenge of finding an initial feasible sectionalizing plan.  To find an initial feasible sectionalizing plan on thousands of buses we next used the randomized approach discussed in Section 4.
}

\subsection{{Solving large scale power systems using the randomized approach}}\label{s:compExp}
In Section~\ref{sec:caseStudy}, the small to medium scale instances IEEE-30, IEEE-118 and {IEEE-300} were solved to optimality within 10 minutes of computational time. The medium scale power systems SC-500 was solved to near optimality in 10 minutes (303 seconds) of computational response time. Therefore, our formulation and bounding approaches work well for small to medium scale power systems. 

{
These approaches, however, do not easily extend to very large power systems because of the large amount of time taken to find an initial feasible sectionalizing plan, which is required by our bounding approach. Thus, we apply the randomized approach discussed in Section 4 to solve these problems. a probabilistic method to find an initial plan on large scale problems. We perform multiple parallel iterations (or runs), with each iteration first finding an initial feasible sectionalizing plan (the \emph{initial plan}), which is then improved by the local search method in the remaining computational response time to obtain the \emph{final plan} of the run. 
}

\begin{table}[hbt!]
\TABLE
{Computational performance of randomized approach on large problems in 10 minutes\label{tab:relax}}{
\centering
\scalebox{1}{
\begin{tabular}{|cc|c|c|c|c|r|r|}
\hline
\up\down & & $|J|$ &Upper bound&Lower bound& Comp. Time & Feasible & LS Improve\\
\hline
\up\down &PEG-1354& 12 &22&20& 224 sec. & 31 runs & 29 runs \\
\hline
\up\down &RTE-1888& 12 &24&20& 340 sec. & 30 runs & 13 runs\\
\hline
\up\down &RTE-1951& 12 &28&22& 496 sec. & 10 runs & 5 runs \\
\hline
\up\down &PEG-2383& 12 &28&23& 573 sec. & 21 runs & 0 runs \\
\hline
\end{tabular}
}
}

\end{table}
\begin{figure}[hbt!]
        \centering
        \begin{subfigure}[b]{0.475\textwidth}
            \centering
            \includegraphics[width=\textwidth]{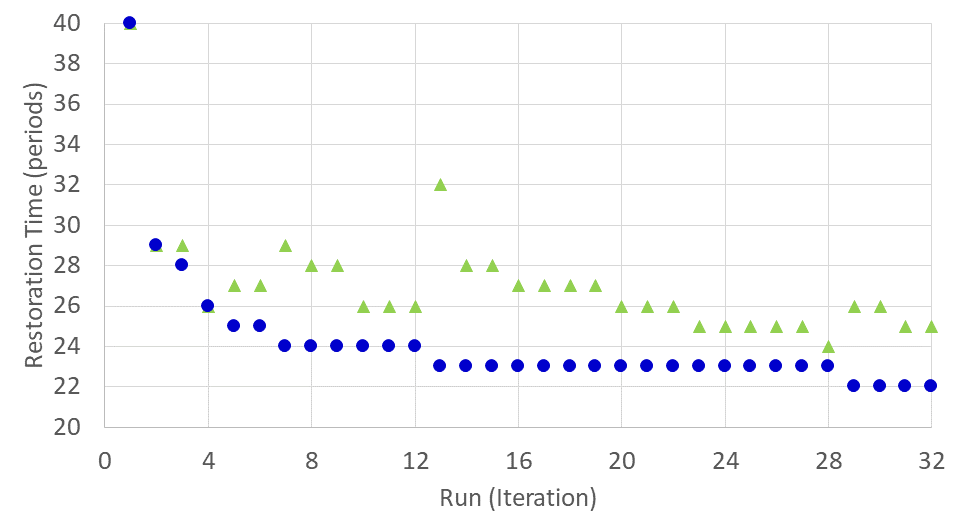}
            \caption{PEG-1354: $RT \leq 22$ periods ($T_{\mathrm{LOW}}=20$)}    
            \label{IEEE-1354}
        \end{subfigure}
        \hfill
        \begin{subfigure}[b]{0.475\textwidth}  
            \centering 
            \includegraphics[width=\textwidth]{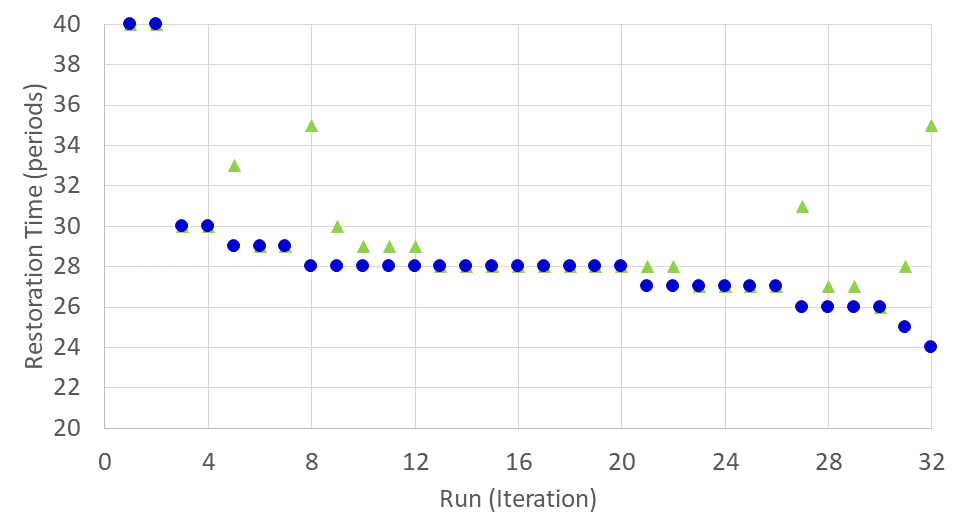}
            \caption{RTE-1888: $RT \leq 24$ periods ($T_{\mathrm{LOW}}=20$)}    
            \label{IEEE-1888}
        \end{subfigure}
        \vskip\baselineskip
        \begin{subfigure}[b]{0.475\textwidth}   
            \centering 
            \includegraphics[width=\textwidth]{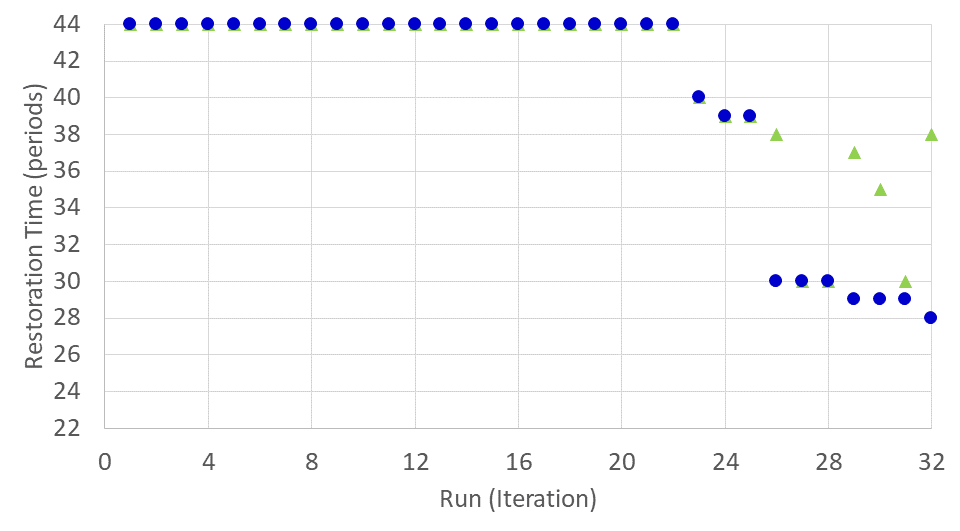}
            \caption{RTE-1951: $RT \leq 28$ periods ($T_{\mathrm{LOW}}=22$)}    
            \label{IEEE-1951}
        \end{subfigure}
        \hfill
        \begin{subfigure}[b]{0.475\textwidth}   
            \centering 
            \includegraphics[width=\textwidth]{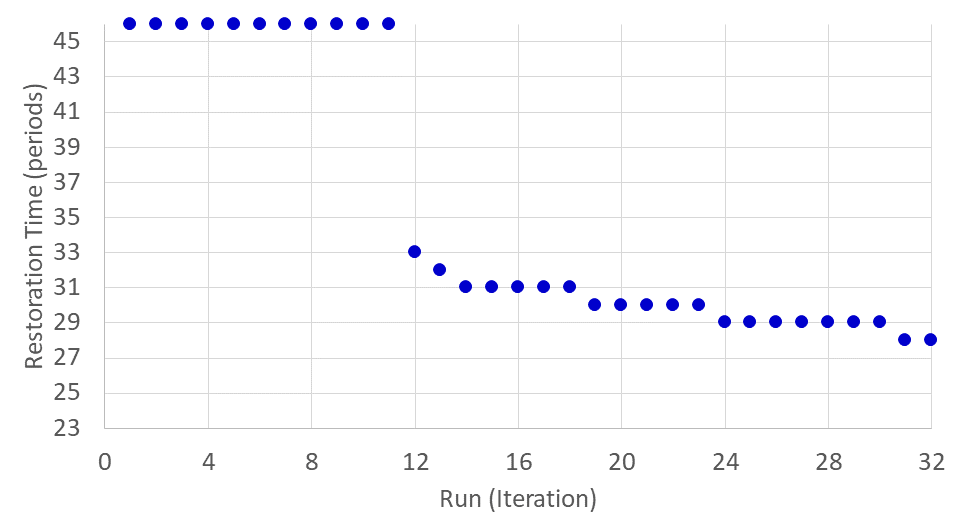}
            \caption{PEG-2383: $RT \leq 28$ periods ($T_{\mathrm{LOW}}=23$)}    
            \label{IEEE-2383}
        \end{subfigure}
            \caption{Results for each of 32 runs (sorted in decreasing order of final solution value) for 10 minutes of computational time}    
            \label{32-10}
\end{figure}

{We perform 32 independent runs (or iterations) in parallel for 10 minutes of computational response time and select the best solution found over all runs. Each run first repeats Algorithm~\ref{a:arr} followed by Algorithm~\ref{a:evaluator} until an initial feasible sectionalizing plan is found. Then it improves the initial plan to the final plan by the local search method.
Each run uses eight threads, and the 32 runs are solved in parallel by two supercomputing nodes, each of which has 128 threads. Our empirical results show that we obtain high quality solutions for all large problems within this time.}

We test the randomized approach on large scale power systems of over a thousand buses: PEG-1354, RTE-1888, RTE-1951, and PEG-2383. We assume the 12 highest degree nodes to be the BS nodes. We assign  NBS generators, critical loads and transshipment nodes to the large scale power systems in the same manner as we did for IEEE-300 and SC-500 in Section~\ref{sec:simulation}.

{
The computational results are summarized in Table~\ref{tab:relax} and Figure~\ref{32-10}. In Table~\ref{tab:relax}, Upper (Lower) bound refers to the best upper (lower) bound found in the 32 runs within 10 minutes (600 seconds). Feasible refers to the number of runs (out of 32) where an initial feasible solution is found. Comp. Time refers to the average computational time needed to find the initial feasible solution. LS Improve refers to the number of runs where local search improved the initial feasible solution within 600 seconds of total run time.}

{
Observe that for PEG-1354, 31 out of 32 runs found an initial feasible sectionalizing plan. Finding a feasible plan took 224 seconds on average across the 31 runs. 29 of these 31 runs then improved the initial plan using local search to a better final plan within the allocated total of 600 seconds. In other words, an average of 224 seconds were spent finding a feasible solution and then 376 seconds were spent on local search. The details of the initial feasible solution and the improved solution after local search for each of the 32 runs for PEG-1354 are shown in Figure 7(a). Observe that for PEG-1354, 4 of the 32 runs obtained a solution of 22, the best available upper bound. }

{
For PEG-2383, 21 of 32 runs obtained an initial feasible solution. Finding an initial feasible solution took 573 seconds on average, leaving hardly any time for local search. As a result, local search could not improve the initial solution on any of the 21 runs.  The best of the 21 initial feasible solutions, however, turns out to have a reasonably good upper bound as shown in Table 2 and Figure 7(d).
}

\begin{figure}
        \centering
        \begin{subfigure}[b]{0.475\textwidth}
            \centering
            \includegraphics[width=\textwidth]{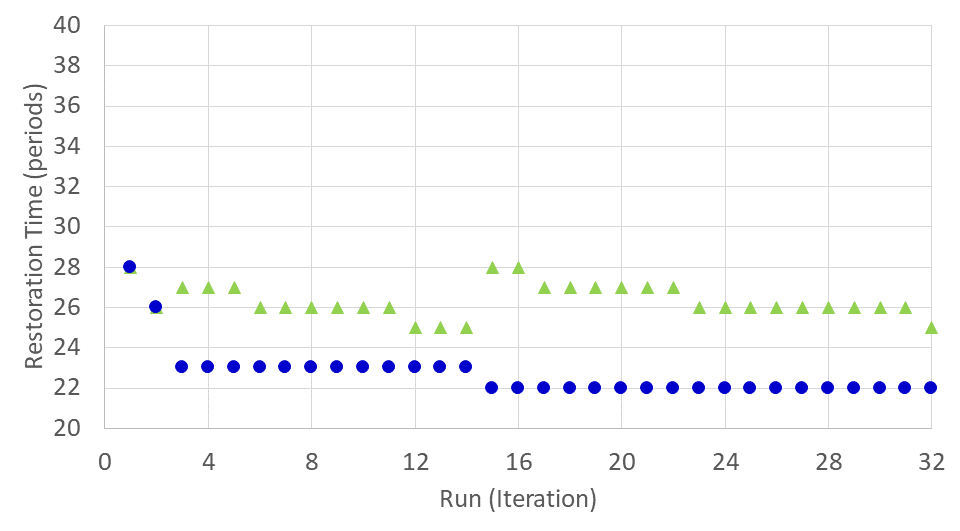}
            \caption{PEG-1354: $RT \leq 22$ periods ($T_{\mathrm{LOW}}=20$)}    
        \end{subfigure}
        \hfill
        \begin{subfigure}[b]{0.475\textwidth}  
            \centering 
            \includegraphics[width=\textwidth]{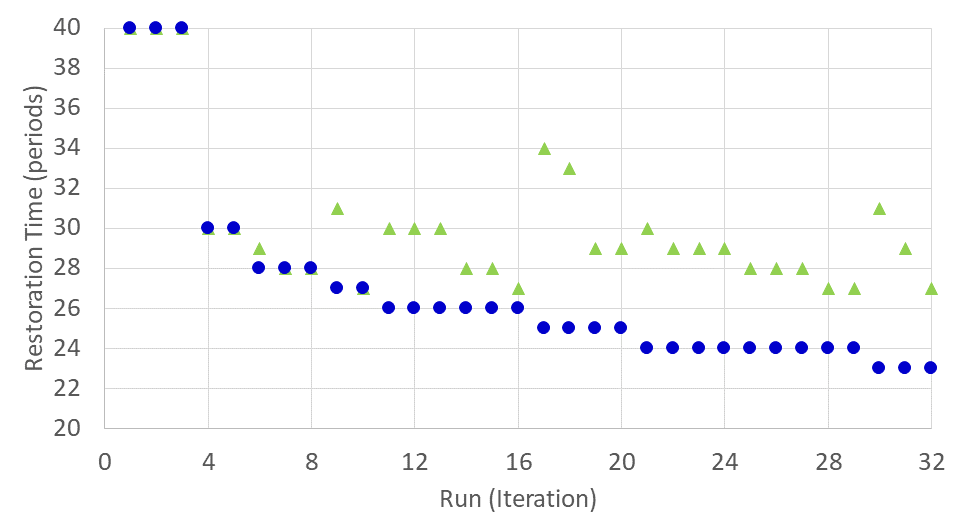}
            \caption{RTE-1888: $RT \leq 23$ periods ($T_{\mathrm{LOW}}=20$)}    
        \end{subfigure}
        \vskip\baselineskip
        \begin{subfigure}[b]{0.475\textwidth}   
            \centering 
            \includegraphics[width=\textwidth]{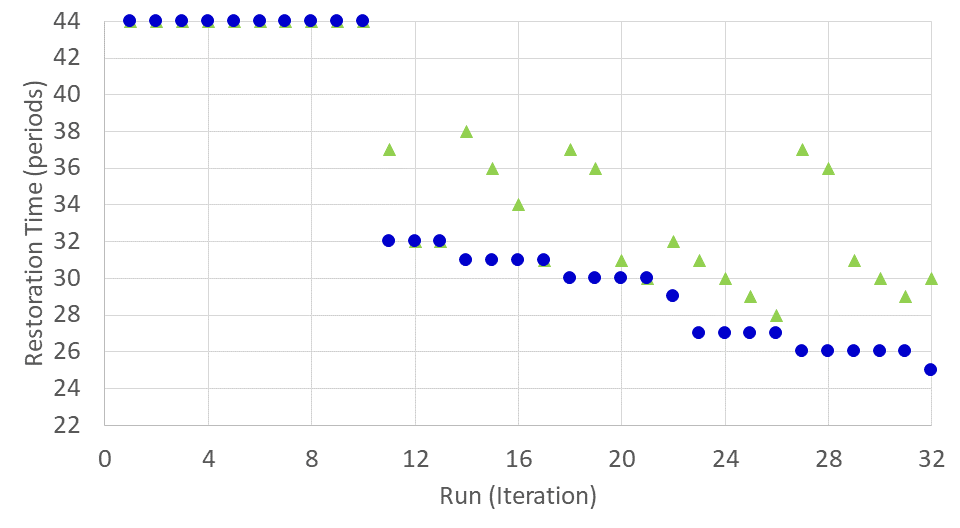}
            \caption{RTE-1951: $RT \leq 25$ periods ($T_{\mathrm{LOW}}=22$)}    
        \end{subfigure}
        \hfill
        \begin{subfigure}[b]{0.475\textwidth}   
            \centering 
            \includegraphics[width=\textwidth]{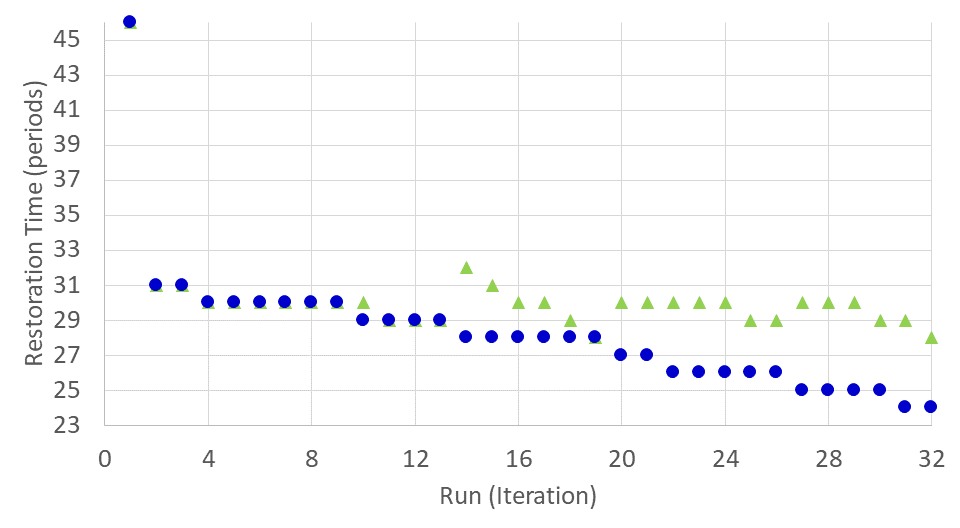}
            \caption{PEG-2383: $RT \leq 24$ periods ($T_{\mathrm{LOW}}=23$)}    
        \end{subfigure}
            \caption{Results for each of 32 runs (sorted in decreasing order of final solution value) for 25 minutes of computational time}    
            \label{32-25}
\end{figure}

{We also perform the same experiments allowing 25 minutes of computational response time. Figure~\ref{32-25} shows the results in this case. Observe that allowing 25 minutes of computation decreases the upper bound for PEG-1888 (from  to 23), RTE-1951 (from 28 to 25), and PEG-2383 (from 28 to 24). Even though the upper bound for PEG-1354 did not decrease, the number of runs achieving the best upper bound of 22 increased from 4 runs in 10 minutes to 18 runs in 25 minutes.}

\section{Conclusion and future work}\label{s:conclusion}
In this paper, we developed an integer programming formulation along with bounding approaches that allow us to solve (within minutes) small and medium sized problems to optimality, while obtaining high quality solutions for very large problems. In future work, our goal is to continue to speed up the solution approaches while also incorporating other elements of the problem into our formulation. Our ultimate goal is to develop a practically usable approach for power system restoration.

\ACKNOWLEDGMENT{This research was supported by the Visiting Faculty Program (VFP) of Argonne National Laboratory and  the U.S. Department of Energy Advanced Grid Modeling Program under Grant DE-OE0000875.
}

\bibliographystyle{nonumber}

\begin{thebibliography}{}

\bibitem[{Adibi \protect\BIBand{} Fink(1994)}]{317561} 
Adibi MM and Fink LH. 1994.
\textquotedblleft Overcoming restoration challenges associated with major power system disturbances\textemdash Restoration from cascading failures.\textquotedblright 
\emph{IEEE Transactions on Power Systems} {9} (1): 22-28. doi:10.1109/59.317561

\bibitem[{Adibi \protect\BIBand{} Fink(2006)}]{Adibi2006} 
Adibi MM and Fink LH. 2006.
\textquotedblleft Overcoming restoration challenges associated with major power system disturbances\textemdash Restoration from cascading failures.\textquotedblright 
\emph{IEEE Power and Energy Magazine} {4} (5): 68--77. https://doi.org/10.1109/MPAE.2006.1687819.

\bibitem[{Adibi et~al.(1987)}]{Adibi1987} 
Adibi M, Clelland P, Fink L, Happ H, Kafka R, Raine J, Scheurer D, and Trefny F. 1987.
\textquotedblleft Power system restoration\textemdash A task force report.\textquotedblright
\emph{IEEE Transactions on Power Systems} {2}: 271--277. https://doi.org/10.1109/TPWRS.1987.4335118.

\bibitem[{Adibi \protect\BIBand{} Kafka(1991)}]{Adibi1991} 
Adibi MM and Kafka RJ. 1991.
\textquotedblleft Power system restoration issues.\textquotedblright
\emph{IEEE Computer Applications in Power} {4} (2): 19--24. https://doi.org/10.1109/67.75871.

\bibitem[{Apollonio et~al.(2008)}]{TreePartition} 
Apollonio N, Lari I, Ricca F, and Simeone B. 2008.
\textquotedblleft Polynomial Algorithms for Partitioning a Tree into Single-Center Subtrees to Minimize Flat Service Costs.\textquotedblright  
\emph{Networks} {51} (1): 78--89. http://doi.org/10.1002/net.20197.

\bibitem[{Birchfield et~al.(2017)}]{Birchfield} 
Birchfield AB, Xu T, Gegner KM, Shetye KS, and Overbye TJ. 2017.
\textquotedblleft Grid Structural Characteristics as Validation Criteria for Synthetic Networks.\textquotedblright
 \emph{IEEE Transactions on Power Systems} {32} (4): 3258--3265. https://doi.org/10.1109/TPWRS.2016.2616385.



\bibitem[{Centolella(2010)}]{Centolella2010} 
Centolella E. 2010.  
\textit{Estimates of the Value of
Uninterrupted Service for the Mid-West
Independent System Operator.} 
https://hepg.hks.harvard.edu/publications/estimates-value-uninterrupted-service-mid-west-independent-system-operator.

\bibitem[{Chopra et~al.(2019)}]{chopraKimshim}
Chopra S, Kim E, and Shim S. 2019. \textquotedblleft Adaptive randomized rounding in the big parsimony problem.\textquotedblright
\emph{Preprint}. http://doi.org/10.13140/RG.2.2.12140.41607.


\bibitem[{Coffrin \protect\BIBand{} Van Hentenryck(2014)}]{pascal2} 
Coffrin C and Van Hentenryck P. 2014.
\textquotedblleft A linear-programming approximation of
ac power flows.\textquotedblright 
\emph{INFORMS Journal on Computing} {26}: 718--734. 

\bibitem[{Coffrin \protect\BIBand{} Van Hentenryck(2015)}]{pascal1} 
Coffrin C and Van Hentenryck P. 2015.
\textquotedblleft Transmission system restoration with co-optimization of repairs, load pickups, and generation dispatch.\textquotedblright 
\emph{Electrical Power and Energy Systems} {72}: 144--154. 

\bibitem[{Conrad et~al.(2006)}]{Conrad2006} 
Conrad SH, LeClaire RJ, O'Reilly GP, and Uzunalioglu H. 2006.
\textquotedblleft Critical national infrastructure reliability modeling and analysis.\textquotedblright 
\emph{Bell Labs Technical Journal} {11} (3): 57--71. http://doi.org/10.1002/bltj.20178.


\bibitem[{Ding et~al.(2013)}]{MLCL2013} 
Ding L, Gonzalez-Longatt F, Wall P, and Terzija V. 2013.
\textquotedblleft Two-step spectral clustering controlled islanding algorithm.\textquotedblright
\emph{IEEE Transactions on Power Systems} {28}: 75--84. 
http://doi.org/10.1109/TPWRS.2012.2197640.

\bibitem[{Ding et~al.(2014)}]{MLCL2014} 
Ding L, Wall P, and Terzija V. 2014.
\textquotedblleft Constrained spectral clustering based controlled islanding.\textquotedblright
\emph{International Journal of Electrical Power \& Energy Systems} {63}: 687--694. http://doi.org/10.1016/j.ijepes.2014.06.016.


\bibitem[{Ferman(2021)}]{texas2021F} 
Ferman M. 2021. 
\textquotedblleft Winter storm could cost Texas more money than any disaster in state history.\textquotedblright
\emph{The Texas Tribune,} February 25, 2021. https://www.texastribune.org/2021/02/25/texas-winter-storm-cost-budget.

\bibitem[{Garey \protect\BIBand{} Johnson(1979)}]{GaJd1979} 
Garey MR, and Johnson DS. 1979.
\emph{Computers and Intractability: A Guide to the Theory of NP-Completeness.}
New York: Freeman.


\bibitem[{Gu \protect\BIBand{} Zhong(2012)}]{gu2012} 
Gu X and Zhong H. 2012.\textquotedblleft 
Optimisation of network reconfiguration based on a two-layer unit-restaring framework for power system restoration.\textquotedblright
\emph{IET Generation, Transmission \& Distribution} {6} (7): 693-700. 

\bibitem[{Jiang et~al.(2017)}]{discretize} 
Jiang Y, Chen S, Liu CC, Sun W, Luo X, Liu S, Bhatt N, Uppalapati S, and Forcum D. 2017.\textquotedblleft 
Blackstart capability planning for power system restoration.\textquotedblright
\emph{International Journal of Electrical Power and Energy Systems} {86}: 127--137. http://doi.org/10.1016/j.ijepes.2016.10.008.

\bibitem[{Kim \protect\BIBand{} Shim(2021)}]{logit} 
Kim SH, and Shim S. 2021.
\textquotedblleft Park-and-ride facility location under nested logit function.\textquotedblright \textit{Preprint}. 
https://arxiv.org/abs/2111.09522.

\bibitem[{Knuth(1969)}]{sphericalUniform} 
Knuth DE. 1969.
\emph{The Art of Computer Programming, vol. 2: Seminumerical
Algorithms}. Reading, MA: Addison-Wesley.


\bibitem[{Lin et~al.(2015)}]{lin2015} 
Liu WJ, Lin ZZ, Wen FS, Chung CY, Xue Y, and Ledwich G. 
2015.
\textquotedblleft Sectionalizing strategies for minimizing outage durations of critical loads in parallel power system restoration with bi-level programming.\textquotedblright
\emph{International Journal of Electrical Power \& Energy Systems} {71}: 327-334.

\bibitem[{Lin et~al.(2011)}]{lin2011} 
Lin ZZ, Wen FS, Chung CY, Wong KP, and Zhou H. 
2011.
\textquotedblleft Division
algorithm and interconnection strategy of restoration subsystems
based on complex network theory.\textquotedblright
\emph{IET Generation, Transmission and Distribution} {5} (6): 674-683.

\bibitem[{Lindenmeyer et~al.(2001)}]{LINDENMEYER2001219} 
Lindenmeyer D, Dommel HW, and Adibi MM. 2001.
\textquotedblleft Power system restoration: a bibliographical survey.\textquotedblright
\emph{International Journal of Electrical Power \& Energy Systems} {23} (3): pp.219-227.

\bibitem[{Liu et~al.(2009)}]{liu2009} 
Liu CC, Vittal V, Heydt G, Tomsovic K, and Sun W. 2009.
\textquotedblleft Development and evaluation of system restoration strategies from a blackout.\textquotedblright
\emph{PSERC Publication} pp.8-9.

\bibitem[{Liu et~al.(2016)}]{8939530} 
Liu Y, Fan R, and Terzija V. 2016.
\textquotedblleft Power system restoration: a literature review from 2006 to 2016.\textquotedblright
\emph{Journal of Modern Power Systems and Clean Energy} {4} (3): pp.332-341.
doi:10.1007/s40565-016-0219-2

\bibitem[{Nagata et~al.(2000)}]{nagata2000} 
Nagata N, Hatakeyama S, Yasouka M, and Sasaki H.\textquotedblleft Anefficient method for power distribution system restoration based on mathematical programming and operation strategy.\textquotedblright
\emph{Proceedings of International Conference on Power System Technology (POWERCON)} December 2000: 1545-1550. 

\bibitem[{Nemhauser \protect\BIBand{} Wolsey(1988)}]{NW1988}
Nemhauser GL, and Wolsey LA. 1988.
\emph{Integer and Combinatorial Optimization.}
New York: Wiley.

\bibitem[{Patsakis et~al.(2018)}]{oren} 
Patsakis G, Rajan D, Aravena I, Rios J, and Oren S. 2018.
\textquotedblleft Optimal Black Start Allocation for Power System Restoration.\textquotedblright 
\emph{IEEE Transactions on Power Systems} {33}: 6766--6776. https://doi.org/10.1109/TPWRS.2018.2839610.

\bibitem[{Perez-Guerrero et~al.(2008)}]{perez-guerrero2008a} 
Perez-Guerrero BK, Heydt GT, Jack NJ, Keel BK, and Castelhano AR.\textquotedblleft Optimal restoration of distribution systems using dynamic programming.\textquotedblright
\emph{IEEE Transactions on Power Delivery} {23} (3): 1589-1596.

\bibitem[{Perez-Guerrero \protect\BIBand{} Heydt(2008)}]{perez-guerrero2008b} 
Perez-Guerrero BK and Heydt GT.\textquotedblleft Distribution system restoration via subgradient-based Lagrangian relaxation.\textquotedblright
\emph{IEEE Transactions on Power Systems} {23} (3): 1162-1169.

\bibitem[{Qiu \protect\BIBand{} Li(2017)}]{restoration} 
Qiu F, and Li P. 2017.
\textquotedblleft An integrated approach for power system restoration planning.\textquotedblright
\emph{Proceedings of the IEEE}  105 (7): 1234-1252. https://doi.org/10.1109/JPROC.2017.2696564.

\bibitem[{Qiu et~al.(2017)}]{blackstart} 
Qiu F, Wang J, Chen C, and Tong J. 2017. 
\textquotedblleft Optimal black start resource allocation.\textquotedblright
\emph{IEEE Transactions on Power Systems} 31  (3): 2493-2494. https://doi.org/10.1109/TPWRS.2015.2442918.


\bibitem[{Quir\'{o}s-Tort\'{o}s \protect\BIBand{} Terzija(2013)}]{graph} 
Quir\'{o}s-Tort\'{o}s J, Terzija V. 2013.
\textquotedblleft A graph theory based new approach for power system restoration.\textquotedblright
\emph{POWERTECH}, Jun 2013, pp. 1-6. DOI: 10.1109/PTC.2013.6652108

\bibitem[{Quir\'{o}s-Tort\'{o}s et~al.(2014)}]{MLCL2014Q} 
Quir\'{o}s-Tort\'{o}s J, S\'{a}nchez-Garc\'{i}a R, Brodzki J, Bialek J, Terzija V. 2014. 
\textquotedblleft Constrained Spectral Clustering Based Methodology for International Controlled Islanding of Large-Scale Power Systems.\textquotedblright
\emph{IET Generation, Transmission \& Distribution} {9} (1): 31--42. https://doi.org/10.1049/IET-GTD.2014.0228.

\bibitem[{Quir\'{o}s-Tort\'{o}s et~al.(2014)}]{PPSR} 
Quir\'{o}s-Tort\'{o}s J, Wall P, Ding L, Terzija V. 2014.
\textquotedblleft Determination of sectionalising strategies for parallel power system restoration: A spectral clustering-based methodology.\textquotedblright
\emph{Electric Power Systems Research} {116}: 381-390. http://doi.org/10.1016/j.epsr.2014.07.005.

\bibitem[{Sarmadi et~al.(2011)}]{WAMS} 
Sarmadi SAN, S. Arash and Dobakhshari, Ahmad Salehi and Azizi, Sadegh and Ranjbar, Ali Mohammad. 
Sarmadi SAN, Dobakhshari AS, Azizi S and, Ranjbar AM. 2011.
\textquotedblleft A Sectionalizing Method in Power System Restoration Based on WAMS.\textquotedblright
\emph{IEEE Transactions on Smart Grid} {2} (1): 190-197. doi:10.1109/TSG.2011.2105510

\bibitem[{Sullivan \protect\BIBand{} Malick(2021)}]{texas2021S} 
Sullivan BK, and Malick NS. 2021. 
\textquotedblleft 5 Million Americans Have Lost Power From Texas to North Dakota After Devastating Winter Storm.\textquotedblright
\emph{Time,} February 15, 2021. https://time.com/5939633/texas-power-outage-blackouts/. 

\bibitem[{Sun et~al.(2011)}]{sun} 
Sun W., Liu C., and Zhang L. 2011. 
\textquotedblleft Optimal generator start-up strategy for
bulk power system restoration.\textquotedblright
\emph{IEEE Transactions on Power Systems} 26 (3): 1357-1366. 



\bibitem[Towns et~al.(2014)]{towns2014xsede}
Towns J, Cockerill T, Dahan M, Foster I, Gaither K, Grimshaw A, Hazlewood V. 2014. \textquotedblleft
XSEDE: Accelerating scientific discovery.\textquotedblright
\emph{Computing in Science \& Engineering} {16}: 62--74. https://doi.org/10.1109/MCSE.2014.80.

\bibitem[{Van Hentenryck \protect\BIBand{} Coffrin(2015)}]{pascal} 
Van Hentenryck P, and Coffrin C. 2015.
\textquotedblleft Transmission system repair and restoration.\textquotedblright 
\emph{Mathematical Programming} {151}: 347--373. 

\bibitem[{Wang \protect\BIBand{} Liu(2009)}]{wang2009} 
Wang HT and Liu YT. 2009. 
\textquotedblleft Multi-objective optimization of power
system reconstruction based on NSGA-II.\textquotedblright
\emph{Automation of Electric Power Systems} 33 (23): 14-18. 

\bibitem[{Wang et~al.(2017)}]{data118} 
Wang D, Gu X, Zhou G, Li S, Liang H. 2017.
\textquotedblleft Decision-making optimization of power system extended black-start coordinating unit restoration with load restoration.\textquotedblright
\emph{International Transactions on Electrical Energy Systems} {27}: 1--18. https://doi.org/10.1002/ETEP.2367.

\bibitem[{Wang et~al.(2011)}]{wang2011} 
Wang C, Vittal V, and Sun K. 2011. 
\textquotedblleft OBDD-based sectionalization strategies for parallel power system restoration.\textquotedblright
\emph{IEEE Transactions on Power Systems} 26 (3): 1426-1433. 

\bibitem[{Weber \protect\BIBand{} Stengle(2021)}]{texas2021} 
Weber PJ and Stengle J. 2021.
\textquotedblleft Texas death toll from February storm, outages surpasses 100.\textquotedblright
\emph{AP News,} March 26, 2021. https://apnews.com/article/hypothermia-health-storms-power-outages-texas-ffeb5d49e1b43032ffdc93ea9d7cfa5f.

\bibitem[{Xavier \protect\BIBand{} Qiu(2020)}]{Matpower} 
Xavier AS, Kazachkov AM, and Qiu F. 2021. ANL-CEEESA/UnitCommitment.jl: v0.2.2 (v0.2.2). Zenodo.org. https://doi.org/10.5281/zenodo.5120043.

\bibitem[{Zhang et~al.(2014)}]{zhang2014} 
Zhang C, Lin Z, Wen F, Ledwich G, and Xue Y. 
2014.
\textquotedblleft Two-stage power network reconfiguration strategy considering node importance and restored generation capacity.\textquotedblright
\emph{IET Generation, Transmission and Distribution} {8} (1): 91-103.

\bibitem[{Zhu \protect\BIBand{} Liu(2014)}]{zhu2014} 
Zhu HN and Liu YT. 2014. 
\textquotedblleft Multi-objective optimization of unit
restoration during network reconstruction considering line
restoration sequence.\textquotedblright
\emph{Automation of Electric Power Systems} 38 (16): 53-59. 
doi:10.7500/AEPS20131104013

\end{thebibliography}

\newpage
\begin{APPENDICES}
\setcounter{secnumdepth}{-1}
\section{Appendix: More about IEEE-118}\label{s:details118}
In Section~\ref{sec:IEEE-118}, we assume that the BS generators supply their full capacity of power immediately after a blackout. The capacity of a BS generator is set to be total load minus critical load of the bus, assuming that it supplies the critical load of the bus and supplies other buses with the remaining capacity in case of a large blackout (it is assumed to be installed to supply total load in case of local blackout). Since bus 25 has only an NBS generator, the BS capacity of the bus is assumed to equal the cranking power of the NBS generator (it is assumed to be installed to supply the NBS generator in case of local blackout). 
The optimal solution to the PPSR problem on the IEEE-118 bus system is shown in  Table~\ref{tab:118}.

\begin{table}
\TABLE
{The optimal solution to the PPSR problem on IEEE-118 Bus System\label{tab:118}}{
\centering
\scalebox{1}{
\begin{tabular}{|ll|r|r|l|}
\hline
\up\down &BS &CAP& RT & Buses in the island\\
\hline
\up\down &21 &12.36 & 20.0 & 33, 34, 35, 36, 37, 39, 40, 41, 42, 43, 44, 49, 18, 19, 20, 21\\
\hline
\up\down &22 &8.87 & 1.0 & 22, 23\\
\hline
\up &25 & 32.39& 20.0 & 5, 8, 9, 10, 25, 26, 27, 30, 32, 38, 60, 61, 62, 64, 65, 66, 67,\\
\down &  &  &    & 68, 79, 80, 81, 97, 98, 113, 114, 115, 116\\
\hline
\up\down &28 & 15.05& 19.0 & 1, 2, 3, 4, 6, 7, 11, 12, 13, 14, 15, 16, 17, 117, 28, 29, 31\\
\hline
\up &45 &48.49 & 20.0 & 24, 45, 46, 47, 48, 63, 69, 70, 71, 72, 73, 74, 75, 76, 77, 78,\\
 & & & &82, 83, 84, 85, 86, 87, 88, 89, 90, 91, 92, 93, 94, 95, 96, 99,\\ 
\down& & & &100, 101, 102, 103, 104, 105, 106, 107, 108, 109, 110, 111, 112, 118\\
\hline
\up\down &51 &15.13 & 20.0 & 50, 51, 52, 53, 54, 55, 56, 57, 58, 59\\
\hline
\end{tabular}
}
}
{Note: The BS column shows the  BS generator buses. The CAP column shows the capacity (MW) of the BS generators. RT is the restoration time of the island in periods. The overall restoration time is 20 periods or 1 hour 40 minutes.}
\end{table}

\end{APPENDICES}

\end{document}